\PassOptionsToPackage{table,xcdraw}{xcolor}
\documentclass[12pt,a4paper]{article}
\pdfoutput=1
\usepackage{jheppub}
\usepackage{gensymb}
\DeclareSymbolFont{matha}{OML}{txmi}{m}{it}
\DeclareMathSymbol{\varv}{\mathord}{matha}{118}
\usepackage{amsmath}
\usepackage{amssymb}
\usepackage{graphicx}
\usepackage{subfigure}
\usepackage{pstricks}
\usepackage{bm}
\usepackage{pbox}
\usepackage{placeins}
\usepackage{booktabs}
\usepackage[T1]{fontenc}
\usepackage{footnote}
\usepackage{pdfpages}
\usepackage{hhline}
\usepackage{multirow}
\usepackage{multicol}
\usepackage{enumitem}
\usepackage[toc,page]{appendix}
\allowdisplaybreaks
\usepackage{comment}
\usepackage{float}
\usepackage{slashed}
\usepackage{xcolor}
\usepackage{textcomp}
\usepackage{gensymb}
\usepackage{multirow}
\usepackage[numbers]{natbib}
\usepackage{notoccite}
\usepackage{dsfont}
\usepackage{rotating}
\usepackage{tabularx}
\usepackage[labelfont=bf]{caption} 
\usepackage{soul}
\usepackage{tabu}
\usepackage{tikz-feynman}
\tikzfeynmanset{compat=1.0.0}
\usetikzlibrary{shapes.misc}
\tikzset{cross/.style={cross out, draw=black, minimum size=2*(#1-\pgflinewidth), inner sep=0pt, outer sep=0pt},cross/.default={3pt}}

\FloatBarrier

\usepackage[many]{tcolorbox}

\renewcommand{\thetable}{\Roman{table}}

\renewcommand{\thetable}{\Roman{table}}

\definecolor{MyDarkBlue}{rgb}{0.1, 0.1, 0.8} 
\definecolor{MyLightBlue}{rgb}{0.22,0.51,0.9}
\definecolor{MyGreen}{rgb}{0.0, 0.5, 0.0}
\definecolor{BrickRed}{rgb}{0.8, 0.25, 0.33}

\RequirePackage{hyperref}
\hypersetup{colorlinks, citecolor=BrickRed,linkcolor=blue, urlcolor=MyLightBlue}

\makeatletter
\gdef\@fpheader{}
\makeatother
\begin{document}

\title{\bf Fully Testable Axion Dark Matter within a Minimal \boldmath{$SU(5)$} GUT}
\author[a]{Stefan Antusch,}
\author[b,c]{Ilja Dor\v{s}ner,}
\author[a]{Kevin Hinze,}
\author[a]{and Shaikh Saad}

\affiliation[a]{Department of Physics, University of Basel, Klingelbergstrasse\ 82, CH-4056 Basel, \\Switzerland}
\affiliation[b]{University of Split, Faculty of Electrical Engineering, Mechanical Engineering and\\ Naval Architecture in Split, Ru\dj era Bo\v{s}kovi\'{c}a 32, HR-21000 Split, Croatia}
\affiliation[c]{J.\ Stefan Institute, Jamova 39, P.\ O.\ Box 3000, SI-1001
  Ljubljana, Slovenia}

\emailAdd{stefan.antusch@unibas.ch, dorsner@fesb.hr,  kevin.hinze@unibas.ch, shaikh.saad@unibas.ch}
\abstract{
We present a minimal Grand Unified Theory model, based on $SU(5)$ gauge symmetry and a global $U(1)$ Peccei-Quinn symmetry, that predicts the existence of an ultralight axion dark matter within a narrow mass range of $m_a\in[0.1,\,4.7]$\,neV. This mass window is determined through an interplay between gauge coupling unification constraints, partial proton decay lifetime limits, and the need to reproduce experimentally observed fermion mass spectrum. The entire parameter space of the proposed model will be probed through a synergy between several low-energy experiments that look for proton decay (Hyper-Kamiokande), axion dark matter through axion-photon coupling (ABRACADABRA and DMRadio-GUT), and nucleon electric dipole moments (CASPEr Electric). 
}
\maketitle

\section{Introduction}
The Standard Model (SM) of elementary particle physics has performed exquisitely in explaining a multitude of experimental observations. There are, however, several important questions that evidently require physics beyond the SM in order to be fully addressed. For example, one of the most important discoveries in particle physics is the observation of nonzero neutrino masses, whereas neutrinos are strictly massless within the SM framework. Furthermore, it is well established that approximately $26\%$ of the total energy density of the universe is in the form of the so-called dark matter that cannot be of the SM origin. This is especially puzzling as the stable SM matter only represents about $5\%$ of the energy density of the universe. Also, the strong CP problem --- why the QCD $\theta$ parameter takes the value $10^{-10}$ or less --- is still an open issue within the SM.

It might be that all these issues are related. In fact, the unified gauge theory~\cite{Pati:1973rp,Pati:1974yy,Georgi:1974sy,Georgi:1974yf,Georgi:1974my,Fritzsch:1974nn} formulation of the elementary particle interactions is a very popular and successful tool for tackling the aforementioned shortcomings of the SM. The simplest possible scenario, among various possible choices of the Grand Unified Theory (GUT) groups, is the Georgi-Glashow model~\cite{Georgi:1974sy} that embeds the entire SM gauge group within an $SU(5)$. In that construction, one 5-dimensional and one 10-dimensional representation of $SU(5)$ comprise all the fermions of a single SM family. The $SU(5)$ symmetry is broken down to the SM gauge group when a real Higgs in the adjoint representation acquires a vacuum expectation value (VEV). The SM symmetry is subsequently broken to $SU(3) \times U(1)_\mathrm{em}$ by the VEV of the SM Higgs doublet that resides within a fundamental representation. The Georgi-Glashow model, however, is incomplete since (i) it fails to achieve gauge coupling unification, (ii) it predicts wrong mass relations between down-type quarks and charged leptons,  and (iii) neutrinos remain massless as in the SM. On top of that, the Georgi-Glashow model does not address the strong CP problem, nor does it include a dark matter candidate.

The most compelling new physics resolution of the strong CP problem is given in terms of the Peccei-Quinn (PQ) symmetry~\cite{Peccei:1977hh,Peccei:1977ur}. In the PQ framework, a global $U(1)_\mathrm{PQ}$ symmetry is spontaneously broken by a complex scalar leading to a nearly massless pseudoscalar particle~\cite{Weinberg:1977ma,Wilczek:1977pj,Kim:1979if,Shifman:1979if,Zhitnitsky:1980tq,Dine:1981rt}, namely the ``axion'', which can, in turn, serve as a cold dark matter candidate~\cite{Preskill:1982cy,Abbott:1982af,Dine:1982ah}. Intriguingly, as first shown in Ref.~\cite{Wise:1981ry}, the axion can be embedded within the scalar representation that breaks the GUT symmetry. The model presented in Ref.~\cite{Wise:1981ry} did not, however, address several important GUT issues, such as neutrino mass generation and gauge coupling unification. For a sample of models that pursue this particular approach, but with a more realistic agenda, see Refs.~\cite{Ernst:2018bib,DiLuzio:2018gqe,FileviezPerez:2019fku, FileviezPerez:2019ssf}. See also Ref.~\cite{Agrawal:2022lsp} for a discussion of other light axion-like particles and their connections to GUTs.

Our primary interest in this manuscript is to combine the PQ symmetry with a simple, yet realistic, $SU(5)$ GUT scenario~\cite{Dorsner:2019vgf,Dorsner:2021qwg} and to investigate the main predictions of such a setup. The $SU(5)$ proposal~\cite{Dorsner:2019vgf,Dorsner:2021qwg} in question extends the particle content of the Georgi-Glashow model by a $35$-dimensional Higgs representation and a $15$-dimensional vectorlike fermion representation. Remarkably, within that scenario, the observed mismatch between the down-type quarks and charged leptons is intrinsically connected to the neutrino mass generation. More specifically, the difference between the down-type quark and charged lepton mass matrices is given by a rank-one matrix. This stipulates that the down-type quarks and charged leptons have similar, yet, different masses, in accordance with experimental observations. The neutrino mass matrix, on the other hand, is made out of a sum of two rank-one matrices that are transpose of each other. This, in turn, dictates that one of the neutrinos is strictly a massless particle. Moreover, since the model relates these three rank-one matrices, the neutrino masses consequentially mirror the mismatch between the down-type quark and charged lepton masses and are thus of the normal hierarchy.

We extend the minimal realistic $SU(5)$ proposal~\cite{Dorsner:2019vgf,Dorsner:2021qwg} with a PQ symmetry to address the strong CP problem as well as the origin of dark matter and show that such a simple extension still preserves the most prominent features of the original model. Our detailed study reveals that the proposed setup is highly predictive, and that the entire parameter space of the theory will be fully tested in the near future through a combination of several experimental efforts. These comprise the proton decay experiment Hyper-Kamiokande as well as the axion dark matter experiments ABRACADABRA, DMRadio-GUT, and CASPEr Electric.

The manuscript is organized as follows. In Sec.~\ref{sec:model} we introduce the particle content and symmetries of the model. The details of the PQ symmetry implementation and the nature of the axion dark matter are discussed in detail in Sec.~\ref{sec:Peccei-Quinn}. A numerical study of the model is performed in Sec.~\ref{sec:numerical_results}, where we also present the most relevant experimental predictions. We briefly conclude in Sec.~\ref{sec:conclusions}.

\section{The model}
\label{sec:model}

The model in question comprises $\overline{5}_{F\,i} \equiv F_{\alpha\,i}$, $10_{F\,j} \equiv T^{\alpha\beta}_j=-T^{\beta\alpha}_j$, $\overline{15}_F \equiv \overline{\Sigma}_{\alpha\beta} = \overline{\Sigma}_{\beta \alpha}$, $15_F \equiv \Sigma^{\alpha\beta}$, $5_{H_a} \equiv \Lambda_a^\delta$ ($a=1,2$),  a complex $24_H \equiv \phi^\alpha_\beta$, $35_H\equiv \Phi_{\alpha\beta\gamma}$, and $24_V \equiv \Gamma^\alpha_\beta$, where $H$s, $F$s, and $V$ denote whether a given irreducible representation, i.e., irrep, contains scalars, fermions, or gauge bosons, respectively, $i,j \:(= 1,2,3)$ represent the generation indices, and $\alpha,\beta,\gamma, \delta \:(= 1,\ldots,5)$ are the $SU(5)$ indices. The decomposition of the $SU(5)$ scalar and fermion irreps under the Standard Model (SM) gauge group $SU(3) \times SU(2) \times U(1)$ is presented in Table~\ref{table:content}.
\begin{table}[h]
\begin{center}
\begin{tabular}{| c | c | c | c |}
\hline
$SU(5)$ & $SU(3)\times SU(2) \times U(1)$  & $SU(5)$ & $SU(3)\times SU(2) \times U(1)$ \\
\hline
\hline
 & $\Xi_a\left(1,2,+\frac{1}{2}\right)$ & & $L_i\left(1,2,-\frac{1}{2}\right)$\\
\raisebox{2ex}[0pt]{$5_{H_a} \equiv \Lambda^{\alpha}_a $} & $\omega_a  \left(3,1,-\frac{1}{3}\right)$  & \raisebox{2ex}[0pt]{${\overline{5}_F}_i \equiv F_{\alpha\,i}$} & $d_i^c\left(\overline{3},1,+\frac{1}{3}\right)$\\ 
\hline
 & $\phi_0 \left(1,1,0\right)$ & & $Q_i\left(3,2,+\frac{1}{6}\right)$\\
 & $\phi_1 \left(1,3,0\right)$ & ${10_F}_i \equiv T_i^{\alpha \beta}$ & $u_i^c\left(\overline{3},1,-\frac{2}{3}\right)$\\
$24_H \equiv \phi^\alpha_\beta$ & $\phi_3\left(3,2,-\frac{5}{6}\right)$ &  & $e_i^c \left(1,1,+1\right)$\\\cline{3-4}
 & $\phi_{\overline{3}} \left(\overline{3},2,+\frac{5}{6}\right)$ &  & $\overline{\Sigma}_1(1,3,-1)$\\
  & $\phi_8 \left(8,1,0\right)$ & $\overline{15}_F \equiv \overline{\Sigma}_{\alpha \beta}$ & $\overline{\Sigma}_3\left(\overline{3},2,-\frac{1}{6}\right)$\\\cline{1-2}
  & $\Phi_1  \left(1,4,-\frac{3}{2}\right)$ & & $\overline{\Sigma}_6\left(\overline{6},1,+\frac{2}{3}\right)$\\\cline{3-4}
  & $\Phi_3  \left(\overline{3},3,-\frac{2}{3}\right)$ & & $\Sigma_1\left(1,3,+1\right)$\\
 \raisebox{2ex}[0pt]{$35_H \equiv \Phi_{\alpha \beta \gamma}$} & $\Phi_6  \left(\overline{6},2,+\frac{1}{6}\right)$ & $15_F \equiv \Sigma^{\alpha \beta}$ & $\Sigma_3\left(3,2,+\frac{1}{6}\right)$\\
 & $\Phi_{10}  \left(\overline{10},1,+1\right)$ &  & $\Sigma_6\left(6,1,-\frac{2}{3}\right)$\\
\hline  
\end{tabular}
\end{center}
\caption{Content and nomenclature of the scalar and fermion irreps of the proposal at both the $SU(5)$ and SM levels. $\alpha,\beta,\gamma \:(= 1,\ldots,5)$ are the $SU(5)$ indices, $i(=1,2,3)$ is a generation index, and $a(=1,2)$ refers to two copies of scalars in the fundamental representation.}
\label{table:content}
\end{table}
We will sometimes, for convenience, refer to a given irrep/multiplet by using either its dimensionality with respect to the appropriate gauge group or the associated symbol.

Beside the non-trivial assignment under the Lorentz symmetry, the aforementioned $SU(5)$ irreps carry the PQ $U(1)_\mathrm{PQ}$ charges that are presented in Table~\ref{tab:PQ_charges}.
\begin{table}[h]
    \centering
    \begin{tabular}{|c|c|c|c|c|c|c|c|c|c|}
    \hline
        $SU(5)$ irrep & $\overline{5}_{F\,i}$ & $10_{F\,i}$ & $\overline{15}_F$ & $15_F$ & $5_{H_1}$ & $5_{H_2}$ & $24_H$ & $35_H$ & $24_V$ \\\hline\hline
        $U(1)_\mathrm{PQ}$ charge & $-\frac{1}{2}$ & $-\frac{1}{2}$ & $-\frac{1}{2}$ & $-\frac{1}{2}$ & $-1$ &  $+1$ & $+1$ & $-1$ & $0$\\
    \hline
    \end{tabular}
    \caption{$U(1)_\mathrm{PQ}$ charge assignment of the model. $H$, $F$, and $V$ subscripts denote scalar, fermion, or gauge boson $SU(5)$ irreps, respectively, while $i= 1,2,3$.}
    \label{tab:PQ_charges}
\end{table}

Before we write down and discuss relevant parts of the model Lagrangian, we briefly justify the proposed particle content. 
\begin{itemize}
    \item $24_H$ breaks the $SU(5) \times U(1)_\mathrm{PQ}$ symmetry. It furthermore provides axion dark matter (DM), helps to generate unification of the SM gauge coupling constants, and facilitates a process of creation of the experimentally observed mismatch between the down-type quark and charged lepton masses. 
    \item $5_{H_1}$ and $5_{H_2}$ jointly break the SM gauge symmetry down to $SU(3) \times U(1)_\mathrm{em}$. $5_{H_2}$ also provides the up-type quark masses through its vacuum expectation value (VEV), whereas $5_{H_1}$ and $5_{H_2}$ together play an indispensable role in three different mechanisms that create phenomenologically viable masses for the down-type quarks, charged leptons, and neutrinos. 
    \item $35_H$ is essential for neutrino mass generation. It also helps to provide the gauge coupling unification at scales compatible with the existing limits on partial proton decay lifetimes. 
    \item $\overline{15}_F$ and $15_F$ participate in the neutrino mass generation mechanism. In addition to that, these $SU(5)$ irreps are, together with $24_H$ and $5_{H_1}$, instrumental in addressing the observed mismatch between the down-type quark and charged lepton masses.
\end{itemize}

\subsection{Scalar sector}
There are several parts of the scalar sector of the model that need to be discussed in detail. The $SU(5) \times U(1)_\mathrm{PQ}$ symmetry breaking is due to
\begin{align}
\label{eq:part_a}
\mathcal{L}&\supset
-\mu^2\phi^{\ast \beta}_\alpha\phi^\alpha_\beta+  \xi_1   (\phi^{\ast \beta}_\alpha\phi^\alpha_\beta)^2
+ \xi_2\phi^{\ast \beta}_\alpha\phi^\alpha_\gamma\phi^{\ast \gamma}_\delta\phi^\delta_\beta
+\xi_3\phi^{\ast \beta}_\alpha\phi^\delta_\gamma\phi^{\ast \alpha}_\beta\phi^\gamma_\delta
+\xi_4\phi^{\ast \beta}_\alpha\phi^\delta_\gamma\phi^{\ast \alpha}_\delta\phi^\gamma_\beta. 
\end{align}
The VEV of $\phi^\alpha_\beta$ that does the $SU(5)$ symmetry breaking reads
\begin{align}
\langle \phi \rangle=\frac{v_\phi}{\sqrt{15}}  \textrm{diag}(-1,-1,-1,3/2,3/2), 
\label{eq:VEV_GUT}
\end{align}
where we assume that the VEV of the electrically neutral component of the $SU(2)$ triplet $\phi_1 (\in 24_H)$ is negligible.   The squares of masses of multiplets in $24_H$, as generated via Eqs.~\eqref{eq:part_a} and \eqref{eq:VEV_GUT}, are  
\begin{align}
&M^2_{\phi_0^\textrm{Re}}=\frac{1}{15} \left(30 \xi _1+7 \xi _2+30 \xi _3+7 \xi _4\right) v_{\phi }^2\equiv m^2_1, \label{massEQ1}
\\
&M_{\phi_0^\textrm{Im}}=0,
\\
&M^2_{\phi_1^\textrm{Re}}=\frac{2}{3} \left(\xi _2+\xi _4\right) v_{\phi }^2 \equiv m^2_3, \label{MASS1}
\\
&M^2_{\phi_1^\textrm{Im}}=\frac{1}{15} \left(\xi _2-30\xi_3+\xi _4\right) v_{\phi }^2,
\\
&M^2_{\phi_8^\textrm{Re}}=\frac{1}{6} \left(\xi _2+\xi _4\right) v_{\phi }^2,
\\
&M^2_{\phi_8^\textrm{Im}}=-\frac{1}{10} \left(\xi _2+20 \xi _3+\xi _4\right) v_{\phi }^2 \equiv m^2_8, \label{MASS2}
\\
&M^2_{\phi_3^\textrm{Re}}=M^2_{\phi_{\overline{3}}^\textrm{Re}}=0,
\\
&M^2_{\phi_3^\textrm{Im}}=M^2_{\phi_{\overline{3}}^\textrm{Im}}=\frac{1}{30} \left(12 \xi _2-60 \xi _3-13 \xi _4\right) v_{\phi }^2 \equiv m^2_{5/6}. \label{massEQ2}
\end{align}
These results are summarized in Table~\ref{tab:24_masses} for convenience. We again emphasize that $24_H$ also breaks the PQ symmetry while we currently discuss solely the $SU(5)$ symmetry breaking. (Hence the omission of an overall phase in Eq.~\eqref{eq:VEV_GUT}. The exact role of that phase will be discussed in Sec.~\ref{sec:Peccei-Quinn}.)
\begin{table}[h]
    \centering
    \begin{tabular}{|c|c|c|}
    \hline
        multiplet & real part mass-squared & imaginary part mass-squared\\
        \hline\hline
        $\phi_0\left(1,1,0\right)$ & $m^2_1$ & $0$\\ 
        $\phi_1\left(1,3,0\right)$ & $m^2_3$ & $\frac{1}{4}m^2_3+m^2_8$\\
        $\phi_8\left(8,1,0\right)$ & $\frac{1}{4}m^2_3$ & $m^2_8$\\
        $\phi_3\left(3,2,-\frac{5}{6}\right)$ & $0$ & $m^2_{5/6}$ \\
        $\phi_{\overline{3}} \left(\overline{3},2,+\frac{5}{6}\right)$ & $0$ & $m^2_{5/6}$\\
    \hline
    \end{tabular}
    \caption{Mass-squared spectrum of a complex irrep $24_H \equiv \phi$. }
    \label{tab:24_masses}
\end{table}

The potential given by Eq.~\eqref{eq:part_a} dictates that the imaginary part of $\phi_0 (\in 24_H)$ is massless. In fact, the axion is mostly composed of that particular state, as we show later on. The real components of $\phi_3 (\in 24_H)$ and $\phi_{\overline{3}} (\in 24_H)$, on the other hand, provide the necessary degrees of freedom for the proton decay mediating gauge bosons in $24_V$ to obtain a mass $M_\mathrm{GUT}$, where 
\begin{equation}
M^2_\mathrm{GUT}=\frac{5 \pi}{6} \alpha_\mathrm{GUT} v^2_{\phi}.
\end{equation}
Here, $M_\mathrm{GUT}$ is also the scale of gauge coupling unification, and $\alpha_\mathrm{GUT}$ is the corresponding $SU(5)$ gauge coupling.

The scalar fields in the fundamental irreps of $SU(5)$ couple via
\begin{align}
\mathcal{L}\supset    
\sum_{a=1}^2\bigg\{
-\frac{1}{2}\mu^2_{\Lambda_a} \Lambda^{\dagger}_a \Lambda_a
+\gamma_{\Lambda_a} \left(\Lambda^{\dagger}_a \Lambda_a\right)^2
\bigg\}
+\zeta_1 \left(\Lambda^{\dagger}_1 \Lambda_1\right) \left(\Lambda^{\dagger}_2 \Lambda_2\right)
+\zeta_2 \left(\Lambda^{\dagger}_1 \Lambda_2\right) \left(\Lambda^{\dagger}_2 \Lambda_1\right)\;,
\end{align}
where we suppress $SU(5)$ indices. The doublet-triplet spitting, i.e., breaking of the mass degeneracy between $\Xi_a$ and $\omega_a$ multiplets, is accomplished via the following additional terms in the scalar potential:
\begin{align}
\mathcal{L}&\supset
\sum_{a=1}^2\bigg\{
\lambda_{\Lambda_a} \Lambda^{\dagger}_a \Lambda_a \phi^\dagger \phi
+
\Lambda^{\dagger}_a\left( \alpha_{\Lambda_a} \phi^\dagger \phi+\beta_{\Lambda_a}  \phi\phi^\dagger  \right) \Lambda_a
\bigg\}
\nonumber\\&
+
\bigg\{
\kappa_1 \Lambda^{\dagger}_2\phi^2\Lambda_1 + \kappa_2 \left(\Lambda^{\dagger}_2\Lambda_1\right)   \phi^2+\mathrm{h.c.}
\bigg\}. \label{pot5524}
\end{align}
The mass-squared matrices of $\omega_a$ and $\Xi_a$ multiplets, in the $\Lambda_1$-$\Lambda_2$ basis, are 
\begin{align}
&M^2_{\omega}= 
\begin{pmatrix}
-\mu^2_{\Lambda_1}+\frac{v^2_\phi}{15}\left( 2\alpha_{\Lambda_1} + 2 \beta_{\Lambda_1}+15 \lambda_{\Lambda_1}\right)
&
v^2_\phi \left( \frac{2}{15}\kappa_1+\kappa_2  \right)
\\
v^2_\phi \left( \frac{2}{15}\kappa_1+\kappa_2  \right)
&
-\mu^2_{\Lambda_2}+\frac{v^2_\phi}{15}\left( 2\alpha_{\Lambda_2} + 2 \beta_{\Lambda_2}+15 \lambda_{\Lambda_2}\right)
\end{pmatrix},
\\
&M^2_{\Xi}= 
\begin{pmatrix}
-\mu^2_{\Lambda_1}+\frac{v^2_\phi}{10}\left( 3\alpha_{\Lambda_1} + 3 \beta_{\Lambda_1}+10 \lambda_{\Lambda_1}\right)
&
v^2_\phi \left( \frac{3}{10}\kappa_1+\kappa_2  \right)
\\
v^2_\phi \left( \frac{3}{10}\kappa_1+\kappa_2  \right)
&
-\mu^2_{\Lambda_2}+\frac{v^2_\phi}{10}\left( 3\alpha_{\Lambda_2} + 3 \beta_{\Lambda_2}+10 \lambda_{\Lambda_2}\right)
\end{pmatrix}.
\end{align}
Clearly, the required doublet-triplet splitting can be obtained by an appropriate choice of the model parameters. The linear combinations of $\omega_1$ and $\omega_2$ will consequently yield mass eigenstates we denote $T_1$ and $T_2$ in the rest of the manuscript. Also, $\Xi_1$ and $\Xi_2$ will produce mass eigenstates $H_1$ and $H_2$, where $H_1$ is identified with the SM Higgs with 125\,GeV mass.

Finally, the VEVs of $5_{H_a}$  that break $SU(3) \times SU(2) \times U(1)$
down to $SU(3) \times U(1)_\mathrm{em}$ are $\langle \Lambda_a \rangle= (0\quad0\quad0\quad0\quad v_{\Lambda_a})^T$.

The lepton number conservation is violated via a single term in the Lagrangian that reads
\begin{align}
\mathcal{L} \supset \lambda \Lambda^\alpha_1 \Lambda^{\beta}_2 \Lambda^{\gamma}_2 \Phi_{\alpha \beta \gamma} +\mathrm{h.c.}.
\label{eq:L}
\end{align}
The neutrino masses will thus be directly proportional to the dimensionless parameter $\lambda$ of Eq.~\eqref{eq:L}.

The masses of the SM gauge group multiplets in $35_H$ are determined by the following $SU(5)$ contractions 
\begin{align}
\mathcal{L} &\supset \mu^2_{35}\Phi\Phi^{\ast}+\lambda_0 \left(\Phi\Phi^{\ast}\right)\phi^\ast\phi
+\lambda_1 \Phi_{\alpha\beta\gamma}(\Phi^{\ast})^{\alpha \delta \epsilon}(\phi^\ast)_\delta^{\beta}\phi_\epsilon^{\gamma} +\lambda_2 \Phi_{\alpha\beta \epsilon}(\Phi^{\ast})^{\alpha\beta \delta}(\phi^\ast)_\gamma^\epsilon\phi_\delta^\gamma  \,.
\label{eq:potential_35}
\end{align}
The contractions of Eq.~\eqref{eq:potential_35} yield 
\begin{align}
&M^2_{\Phi_1}=\mu^2_{35}+ v^2_\phi \left( \frac{\lambda_0}{2} +\frac{3\lambda_1}{20} +\frac{3\lambda_2}{20} \right),
\\
&M^2_{\Phi_3}=\mu^2_{35}+ v^2_\phi \left( \frac{\lambda_0}{2} -\frac{\lambda_1}{60} +\frac{11\lambda_2}{90} \right),
\\
&M^2_{\Phi_6}=\mu^2_{35}+ v^2_\phi \left( \frac{\lambda_0}{2} -\frac{2\lambda_1}{45} +\frac{17\lambda_2}{180} \right),
\\
&M^2_{\Phi_{10}}=\mu^2_{35}+ v^2_\phi \left( \frac{\lambda_0}{2} +\frac{\lambda_1}{15} +\frac{1\lambda_2}{15} \right).
\end{align}
These, in turn, produce 
a single mass-squared relation that reads 
\begin{align}
\label{eq:mass_squared_relation}
&M^2_{\Phi_{10}}=M^2_{\Phi_1}-3M^2_{\Phi_3}+3 M^2_{\Phi_6}.
\end{align}

The mass spectrum given in Table~\ref{tab:24_masses} and the mass relation presented in Eq.~\eqref{eq:mass_squared_relation} are necessary input for the gauge coupling unification analysis.

\subsection{Fermion sector}\label{sec:fermion_sector}
The Yukawa sector of the model is
\begin{align}
\mathcal{L}&\supset 
Y^u_{ij}\;10_{F\,i} 10_{F\,j}5_{H_2}
+Y^d_{ij}\;10_{F\,i}\overline{5}_{F\,j} 5^\ast_{H_1}
+Y^{a}_{i}\;15_F \overline{5}_{F\,i} 5^\ast_{H_1}
\nonumber \\ &
+Y^{b}_{i}\;\overline{15}_F \overline{5}_{F\,i} 35^\ast_H
 +Y^{c}_{i}\; 10_{F\,i} \overline{15}_F 24_H
+y\; \overline{15}_F 15_F 24_H
+\mathrm{h.c.},  
\label{eq:yukawa}
\end{align}
where the PQ charge assignment of Table~\ref{tab:PQ_charges} and the $SU(5)$ indices are all implicitly understood. The Yukawa matrix elements of the model are $Y^u_{ij} \equiv Y^u_{ji}$, $Y^d_{ij} = Y^{d*}_{ij} \equiv \delta_{ij} Y^d_{i}$, $Y^{a}_{i}$, $Y^{b}_{i}$, $Y^{c}_{i}$, and $y$, where we have used the freedom to rotate irreps in the $SU(5)$ group space to reach this particular Yukawa coupling basis. The model accordingly has nineteen real parameters and fifteen phases in the Yukawa sector to accommodate all of the masses and mixing parameters of the SM fermions as well as the masses of fermions in the $\overline{15}_F$-$15_F$ vectorlike pair.

The PQ charge assignment forbids a bare-mass term for the $\overline{15}_F$-$15_F$ pair. The masses of the associated SM gauge group multiplets are thus generated solely through the last term of Eq.~\eqref{eq:yukawa}, which reads
\begin{align}
\mathcal{L}&\supset  
\frac{y v_\phi}{\sqrt{15}} \left(\frac{3}{2}\overline\Sigma_1\Sigma_1+ \frac{1} {4}\overline\Sigma_3\Sigma_3 -\overline\Sigma_6\Sigma_6 \right)+\mathrm{h.c.}, 
\end{align}
where the overall phase of $24_H$, once again, is not shown for simplicity. We subsequently define 
\begin{align}
\label{eq:mass_relation_a}
&M_{\Sigma_1}=\frac{y}{2} \sqrt{\frac{3}{5}} v_\phi\,,\\
\label{eq:mass_relation_b}
&M_{\Sigma_3}=\frac {y} {4\sqrt {15}} v_\phi\,,\\
\label{eq:mass_relation_c}
&M_{\Sigma_6}=-\frac{y}{\sqrt{15}} v_\phi.
\end{align}
It is important to point out that, apart from different Clebsch-Gordan coefficients, all submultiplets within $15_F$ have a common mass scale. 
(Even though $\Sigma_1$ and $\Sigma_3$ mix with the fermions in $\overline{5}_{F\,i}$ and $10_{F\,i}$, this does not affect equalities in Eqs.~\eqref{eq:mass_relation_a} and \eqref{eq:mass_relation_b}.) We will show, later on, that the product
$y v_\phi$ is rather constrained by a requirement for the model to simultaneously generate large enough unification and neutrino mass scales. 

The masses of the SM fermions are obtained after the breaking of the SM gauge group down to $SU(3) \times U(1)_\mathrm{em}$ as follows. The down-type quark sector $4\times 4$ mass matrix can be written as
\begin{align}
M_D=\begin{pmatrix}
v_{\Lambda_1} Y^d & v^\prime_\phi Y^c\\
v_{\Lambda_1} Y^a & M_{\Sigma_3}
\end{pmatrix},
\end{align}
where we introduce $v^\prime_\phi=-\frac{1}{4}\sqrt{\frac{5}{3}}v_\phi$. This matrix can be transformed into a block-diagonal form comprising a $3 \times 3$ part denoted $M_d$ and a mass parameter $M_H$ as follows
\begin{align}
&X M_D Y^\dagger=\begin{pmatrix}
M_d & 0\\
0 & M_H
\end{pmatrix},
\end{align}
where unitary matrices $X$ and $Y$ take the form 
\begin{align}
&X\sim  \begin{pmatrix}
\left(\mathds{1}+\frac{v^{\prime 2}_\phi}{M^2_{\Sigma_3}} Y^c{Y^c}^\dagger \right)^{-1/2} &-\left(\mathds{1}+\frac{v^{\prime 2}_\phi}{M^2_{\Sigma_3}} Y^c{Y^c}^\dagger \right)^{-1/2} \frac{v^\prime_\phi}{M_{\Sigma_3}} Y^c\\
\frac{v^\prime_\phi{Y^c}^\dagger}{M_H}&\frac{M_{\Sigma_3}}{M_H}
\end{pmatrix},
\\&
Y\sim  \begin{pmatrix}
\mathds{1}&-\frac{v_{\Lambda_1} v^\prime_\phi}{M_H^2}  ({Y^d}^\dagger Y^c+\frac{M_{\Sigma_3}}{v^\prime_\phi} {Y^a}^\dagger)\\
\frac{v_{\Lambda_1} v^\prime_\phi}{M_H^2}  ({Y^c}^\dagger Y^d+\frac{M_{\Sigma_3}}{v^\prime_\phi} Y^a)&1
\end{pmatrix},
\end{align}
with
\begin{align}
\label{eq:down}
&M_d\sim \left(\mathds{1}+\frac{v^{\prime 2}_\phi}{M^2_{\Sigma_3}} Y^c{Y^c}^\dagger \right)^{-1/2} \left( v_{\Lambda_1} Y^d- \frac{v_{\Lambda_1} v^\prime_\phi}{M_{\Sigma_3}}Y^cY^a \right),
\\&
M_H=\sqrt{M_{\Sigma_3}^2+{v^\prime_\phi}^2{Y^c}^\dagger Y^c}\approx M_{\Sigma_3}.
\end{align}
Here, $\mathds{1}=\textrm{diag}(1,1,1)$ while $Y^c$, $Y^a$, and $Y^d$ are Yukawa matrices that are featured in Eq.~\eqref{eq:yukawa}. It is clear from Eq.~\eqref{eq:down} that the down-type quark mass matrix $M_d$ is generated through the VEV of $5_{H_1}$ and the mixing between fields in $\overline{5}_{F\,i}$,  $10_{F\,i}$, $\overline{15}_F$, and $15_F$. This is possible due to the fact that $\overline{\Sigma}_3 \in \overline{15}_F$ and $Q_i \in 10_{F\,i}$ transform in the exact same way under the SM gauge group~\cite{Oshimo:2009ia}.

The charged fermion mass matrices of the model can be succinctly written as
\begin{align}
&M_u=\left(\mathds{1}+\delta^2\;Y^c{Y^c}^{\dagger} \right)^{-\frac{1}{2}} 8 v_{\Lambda_2} Y^u, \label{eq:massu}
\\
&M_d=\left(\mathds{1}+\delta^2\;Y^c{Y^c}^{\dagger} \right)^{-\frac{1}{2}} v_{\Lambda_1} \left( Y^d + \delta\; Y^cY^a  \right),  
\label{eq:massd}
\\
&M_e=v_{\Lambda_1} {Y^d}
\label{eq:masse},
\end{align}
where $\delta=-v^\prime_\phi/M_{\Sigma_3}$ and $v^2_{\Lambda_1}+v^2_{\Lambda_2}=v^2$ with $v=174$\,GeV. We note the two most prominent features of the charged fermion sector. First, $M_u$ can be treated as a symmetric matrix in the flavor space. Second, a mismatch between the charged lepton and down-type quark mass matrices is proportional to a rank-one matrix $Y^c Y^a$. We again stress that we work in the basis where $Y^u_{ij} \equiv Y^u_{ji}$ and $Y^d_{ij} = Y^{d*}_{ij} \equiv \delta_{ij} Y^d_{i}$. This simply means that $v_{\Lambda_1} {Y^d}_i$, where $i=1,2,3$, are the masses of the SM charged leptons.

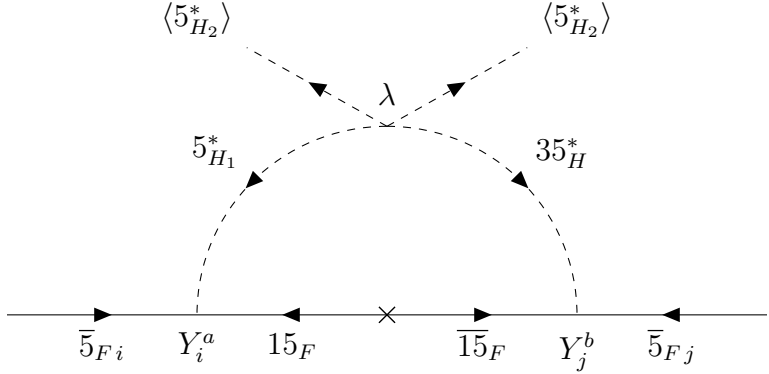
\begin{figure}
    \centering
    \begin{tikzpicture}[inner sep=5pt]
\begin{feynman}
\vertex(a1);
\vertex[right=10cm of a1](a5);
\vertex[right=2.5cm of a1](a2);
\vertex[below=0.5mm of a2](label1){\(Y_i^a\)};
\vertex[right=7.5cm of a1](a3);
\vertex[right=5cm of a1](a4);
\vertex[below=0.5mm of a3](label2){\(Y_j^b\)};
\vertex[above=2.5cm of a4](a6);
\vertex[above=3.5cm of a2](a7){\(\langle 5_{H_2}^{*}\rangle\)};
\vertex[above=3.5cm of a3](a8){\(\langle 5_{H_2}^{*}\rangle\)};
\vertex[above=1mm of a6](label3){\(\lambda\)};
\diagram* {
(a1)--[fermion, edge label'=\(\overline{5}_{F\,i}\)](a2), 
(a6) --[charged scalar, quarter right, edge label'=\(5_{H_1}^*\)](a2),
(a6) --[charged scalar, quarter left, edge label=\(35_H^*\)](a3),
(a4) --[fermion, edge label=\(15_F\),inner sep=7pt, insertion=0](a2),
(a4) --[fermion, edge label'=\(\overline{15}_{F}\)](a3),
(a5)--[fermion, edge label=\(\overline{5}_{F\,j}\)](a3),
(a6) --[charged scalar](a7),
(a6) --[charged scalar](a8),
};
\end{feynman}
\end{tikzpicture}
    \caption{The 1-loop Feynman diagram which is responsible for neutrino mass generation.}
    \label{fig:neutino mass}
\end{figure}

The neutrino mass in this model is generated by utilizing the Yukawa couplings $Y^a$ and $Y^b$ that appear in Eq.~\eqref{eq:yukawa} and the lepton number violating term of Eq.~\eqref{eq:L}. Completion of the neutrino mass loop requires, in addition to the SM fields, the presence of $(1,3,1)+(1,3,-1)(\subset 15_F+\overline{15}_F)$ vectorlike fermions and the scalar quadruplet $(1,4,-3/2)(\subset 35_H)$. The corresponding Feynman diagram illustrating the neutrino mass generation mechanism is shown in Fig.~\ref{fig:neutino mass}. This particular one-loop mechanism to generate neutrino masses has been introduced in Ref.~\cite{Babu:2009aq,Bambhaniya:2013yca}.

The neutrino mass matrix elements $(M_\nu)_{ij}$, at the leading order, read 
\begin{align}
\nonumber
(M_\nu)_{ij}&\approx \frac{\lambda v_{\Lambda_2}^2}{8 \pi^2} (Y^a_iY^b_j+Y^b_iY^a_j) \frac{M_{\Sigma_1}}{M^2_{\Sigma_1}-M^2_{\Phi_1}}
\ln \left( \frac{M^2_{\Sigma_1}}{M^2_{\Phi_1}} \right)
\\&
\equiv m_0 (Y^a_iY^b_j+Y^b_iY^a_j) = (N\;
 \mathrm{diag}(0,m_2,m_3)\;
N^T)_{ij}\,, \label{eq:massnu}
\end{align}
where $m_2$ and $m_3$ are neutrino mass eigenstates and $N$ is a unitary matrix. Note that one of the neutrinos is a strictly  massless particle due to the fact that $M_\nu$ is constructed out of two rank-one matrices with elements $Y^a_iY^b_j$ and $Y^b_iY^a_j$. This is accordingly encoded in the right-hand side of Eq.~\eqref{eq:massnu}. 

Since the charged lepton mass matrix in Eq.~\eqref{eq:masse} is already in a diagonal form, we can write that 
\begin{align}
N= \mathrm{diag}(e^{i \eta^\nu_1},e^{i \eta^\nu_2},e^{i \eta^\nu_3}) V^*_\mathrm{PMNS},
\end{align}
where $V_\mathrm{PMNS}$ is the Pontecorvo-Maki-Nakagawa-Sakata (PMNS) unitary mixing matrix, that is defined as $V_\mathrm{PMNS}=R_{23}U_{13}R_{12}Q$, with $Q=\mathrm{diag}(1,e^{i\beta^\nu},1)$. Here we use the PDG parametrization~\cite{ParticleDataGroup:2022pth} for the $R_{23}$, $U_{13}$, and $R_{12}$ matrices. Note that there is only one Majorana phase $\beta^\nu$ appearing in $Q$ due to the fact that one of the neutrinos is massless.

One especially convenient feature of the neutrino sector is that the matrices $Y^a$ and $Y^b$ can be expressed in terms of the PMNS matrix parameters and phases $\eta^\nu_i$, $i=1,2,3$. Using the parametrization mentioned in Refs.~\cite{Cordero-Carrion:2018xre,Cordero-Carrion:2019qtu} we can write the two Yukawa coupling vectors $Y^a$ and $Y^b$ as  
\begin{equation}
\label{eq:perturbativity}
Y^{a\,T}=\frac{\xi}{\sqrt{2}} \begin{pmatrix}
i\;r_2\;N_{12}+r_3\;N_{13}\\
i\;r_2\;N_{22}+r_3\;N_{23}\\
i\;r_2\;N_{32}+r_3\;N_{33}
\end{pmatrix},\;\;
Y^{b\,T}=\frac{1}{\sqrt{2}\xi} \begin{pmatrix}
-i\;r_2\;N_{12}+r_3\;N_{13}\\
-i\;r_2\;N_{22}+r_3\;N_{23}\\
-i\;r_2\;N_{32}+r_3\;N_{33}
\end{pmatrix}\;,
\end{equation}
where $N_{ij}$ denotes the $ij$-th element of the unitary matrix $N$, $r_2=\sqrt{m_2/m_0}$, and $r_3=\sqrt{m_3/m_0}$. 
Moreover, $\xi$ is a dimensionless scaling parameter that needs to be introduced if one is to accurately scan over all possible phenomenologically viable entries in $Y^a$ and $Y^b$ that accommodate experimental observables in the neutrino sector with utmost certainty. Eq.~\eqref{eq:perturbativity} is applicable solely to the normal neutrino mass hierarchy scenario since that is one of the model predictions, as we will discuss later.
For alternative ways of generating neutrino masses within the $SU(5)$ framework, see, for example, Refs.~\cite{Dorsner:2005fq,Bajc:2006ia,Perez:2007rm,Perez:2016qbo,Dorsner:2017wwn,Kumericki:2017sfc,Saad:2019vjo,Klein:2019jgb,Antusch:2021yqe,Antusch:2022afk}.

\section{Peccei-Quinn symmetry and axion dark matter}
\label{sec:Peccei-Quinn}

We discuss the implementation of the PQ symmetry within our setup and elaborate on the model's main ingredients and experimental detection prospects in the following.   

In the ``invisible axion'' models~\cite{Kim:1979if,Shifman:1979if,Zhitnitsky:1980tq,Dine:1981rt} the PQ symmetry is broken by a scalar field that carries a non-trivial PQ charge, where the scalar is a singlet under the SM. We embed this scalar within the $24$-dimensional Higgs irrep that is charged under the $U(1)_\mathrm{PQ}$ symmetry, as shown in Table~\ref{tab:PQ_charges}. Consequently, our setup unifies the GUT and PQ breaking scales. The VEV of $24_H \equiv \phi^\alpha_\beta$ can be written as~\cite{Wise:1981ry}
\begin{align}
\langle \phi\rangle=\frac{\hat  v_\phi}{\sqrt{2}}  \textrm{diag}\left(\frac{-1}{\sqrt{15}},\frac{-1}{\sqrt{15}},\frac{-1}{\sqrt{15}},\frac{3}{2\sqrt{15}},\frac{3}{2\sqrt{15}}\right)    e^{i a_\phi(x)/\hat  v_\phi}, \;\;\; \hat  v_\phi\equiv \sqrt{2}v_\phi,
\end{align}
where the pseudoscalar part, i.e., field $a_\phi(x)$, essentially remains massless, whereas the radial mode acquires a mass of the order of the GUT scale while the global $U(1)_\mathrm{PQ}$ symmetry is spontaneously broken with order parameter $v_\phi$. To correctly identify the massless axion, one also needs to include all other Higgses that carry PQ charges and participate in symmetry breaking. 

The non-Hermitian operators that are responsible for the breaking of the re-phasing symmetry of the three scalar fields are given by the terms in the second line of Eq.~\eqref{pot5524}. 
The VEVs of neutral components of the $SU(2)$ doublets can be re-written as
\begin{align}
\langle \Lambda_2\rangle= \frac{\hat v_{\Lambda_2}}{\sqrt{2}}e^{i\frac{a_{\Lambda_2}}{\hat v_{\Lambda_2}}},\;\;\;
\langle \Lambda^\ast_1\rangle=
\frac{\hat v_{\Lambda_1}}{\sqrt{2}}e^{i\frac{a_{\Lambda_1}}{\hat v_{\Lambda_1}}},\;\;\;
\hat v_{\Lambda_a}\equiv \sqrt{2} v_{\Lambda_a},
\end{align}
where we take all VEVs to be real, and, as mentioned before, we neglect the VEV of the $SU(2)$ triplet in $24_H$. With these assumptions, the axion field is identified as~\cite{Srednicki:1985xd},
\begin{align}
a=\frac{x_{\Lambda_2}\hat v_{\Lambda_2} a_{\Lambda_2} +x^\ast_{\Lambda_1} \hat v_{\Lambda_1} a_{\Lambda_1} + x_\phi \hat v_\phi a_\phi  }{v_a},\;\;\; v^2_a= x_{\Lambda_2}^2\hat v_{\Lambda_2}^2 +x_{\Lambda_1}^2 \hat v_{\Lambda_1}^2  + x_\phi^2 \hat v_\phi^2, 
\end{align}
where $x_i$ denotes the PQ charge of the corresponding $i$-th scalar (and $x_i^\ast=-x_i$). Since $v_\phi\sim 10^{16}$\,GeV and $v_{\Lambda_a}\sim 10^2$\,GeV, the axion mostly resides in $24_H$ with $a\approx a_\phi$. 

The axion field must also be orthogonal to the Goldstone field eaten up by the $Z$-boson. This translates into the following condition 
\begin{align}\label{eq:tanbeta}
\tan^2\beta=\frac{ v^2_{\Lambda_2} }{v^2_{\Lambda_1}}= \frac{x^\ast_{\Lambda_1}}{x_{\Lambda_2}},    
\end{align}
which, in our benchmark charge assignment, fixes $\tan\beta=1$. Here, we do not present the expression of the SM Higgs mass eigenstate, which can be obtained via the diagonalization of the $4\times 4$ mass matrix of the CP-even states. The heaviest one is expected to reside at the GUT scale, and the lightest one is the SM Higgs boson.  Depending on the chosen hierarchy, the remaining two eigenstates --- one coming from the triplet and the other from the pair of doublets --- can live anywhere in between the electroweak and GUT scales.

Now, performing a field-dependent axial transformation that is anomalous under QCD, the axion can be disentangled from the Yukawa interactions. This transformation generates the effective anomalous interactions of the following types:
\begin{align}
\delta \mathcal{L}_\mathrm{eff}= \frac{\alpha_s}{8\pi}\frac{a}{f_a} G\widetilde G    
+
\left( 
\frac{\alpha_\mathrm{em}}{2\pi f_a} \frac{\mathcal E}{\mathcal N} \right) \frac{a}{4} F\widetilde F  \;.   
\end{align}
Here, $G$ ($F$) is the gluon (photon) field strength tensor, $\widetilde G$ ($\widetilde F$) is its dual, and $f_a$ is the axion decay constant. The effective operator of the form $aG\widetilde G$ is the key to the PQ solution to the strong CP problem. Since these sub-multiplets carry  color and electromagnetic charges, the PQ current has both QCD and electromagnetic anomalies, with the corresponding anomaly coefficients~\cite{DiLuzio:2020wdo},
\begin{align}
&\mathcal N=\sum_{\psi} N_\psi, \;\;\;
\mathcal E=\sum_{\psi} E_\psi\;,
\end{align}
where sums are taken over all fermions, which we generically
 denote by $\psi$. Using well-known formulas, 
\begin{align}
&N_\psi=x_\psi d(I_\psi) T(C_\psi),  
\\
&E_\psi=x_\psi d(C_\psi) d(I_\psi) \left( \frac{1}{12}(d(I_\psi)^2-1) +Y^2_\psi\right),  
\end{align}
we obtain $|\mathcal N|\equiv \mathcal{\hat N} =13/2$ and $|\mathcal E| \equiv \mathcal{\hat E} =52/3$ while the domain-wall number, which is relevant for cosmology, is $N_\mathrm{DW}=2 \mathcal{\hat N} = 13$. Subsequently, we find the axion decay constant to be
\begin{align}
f_a=\frac{v_a}{2 \mathcal{\hat N}}\approx \frac{\hat v_\phi}{2 \mathcal{\hat N}} = \sqrt{\frac{3}{10\pi\alpha_\textrm{GUT}}}\frac{M_\textrm{GUT}}{\mathcal{\hat N}}.   \label{decayconstant}
\end{align}
Since the decay constant is of the order of the GUT scale, i.e., $f_a\sim M_\mathrm{GUT}$, we refer to the axion as the ``GUT axion''. Once strong interactions confine, non-perturbative QCD effects generate a potential that gives rise to a tiny axion mass~\cite{Bardeen:1978nq,GrillidiCortona:2015jxo}
\begin{align}\label{eq:axion_mass}
    m_a=5.7\, \textrm{neV}\; \left( \frac{10^{15}\,\textrm{GeV}}{f_a} \right)   
=5.7\, \textrm{neV}\left(\frac{10^{15}\, \textrm{GeV}}{M_\textrm{GUT}}\right)\mathcal{\hat N}\sqrt{\frac{10\pi\alpha_\textrm{GUT}}{3}}\;.
\end{align}
This shows that the axion mass is predicted if the grand unification scale $M_\textrm{GUT}$ is known. We accordingly compute the predicted range of the GUT scale within our model in Sec.~\ref{sec:numerical_results} by taking into account all relevant constraints.

\begin{figure}[t!]
\centering
\includegraphics[scale=0.9]{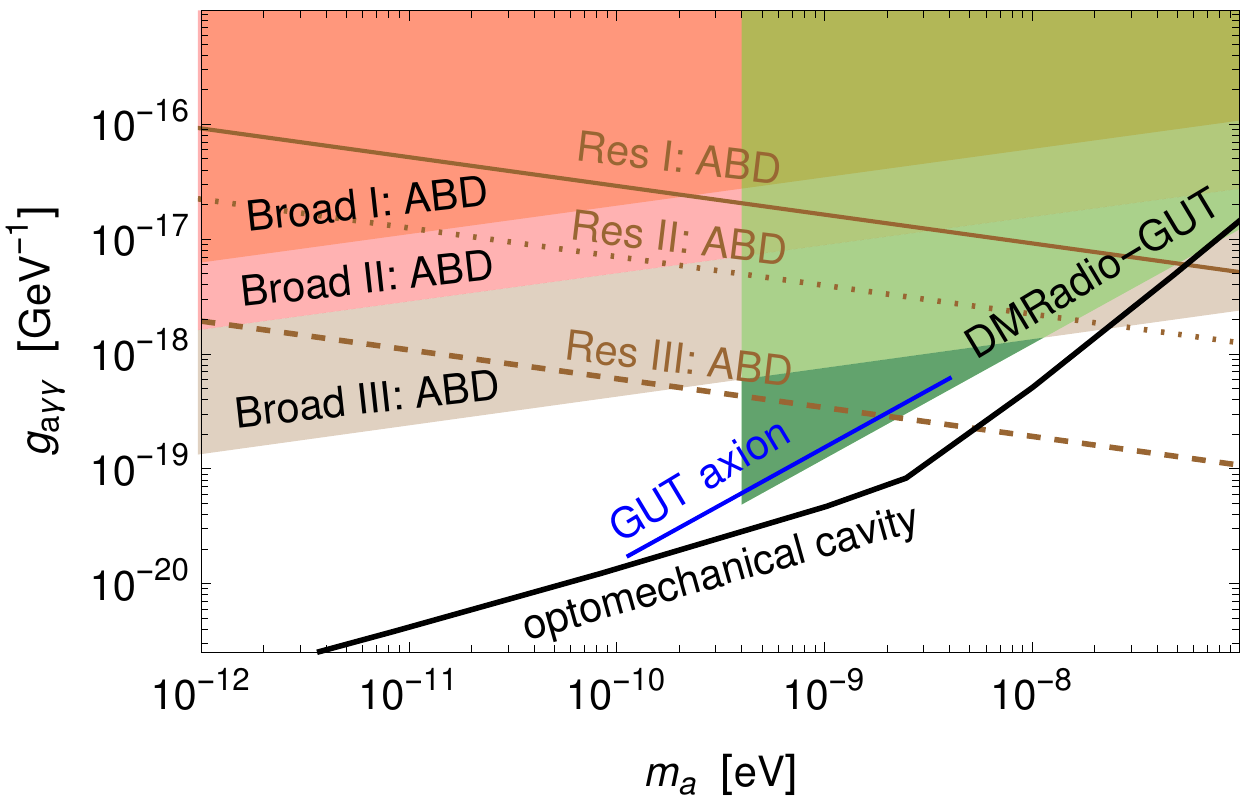}
\caption{  Expected reach in the $m_a$ vs.\ $g_{a\gamma\gamma}$  plane for the broadband (Broad) and resonant (Res) strategies of the ABRACADABRA (ABD) experiment~\cite{Kahn:2016aff}. The blue line (that lies on the QCD line) corresponds to the prediction of our model. The projected 3\,$\sigma$ sensitivity of DMRadio-GUT~\cite{Domcke:2022rgu,DMRadio} is also presented in the green shaded region. Furthermore, the expected theoretical reach using the optomechanical cavity method~\cite{Murgui:2022zvy} is shown with solid black lines. See text for more details.}\label{fig:DM}
\end{figure}
Since the non-observation of proton decay requires the GUT scale to be large, the axion mass is expected to be around the neV scale within our setup. An axion in this mass range is extremely weakly coupled to the SM particles due to an extremely large decay constant. Remarkably, an axion with neV mass can serve as an excellent dark matter candidate and can be searched for efficiently in direct detection experiments~\cite{Adams:2022pbo} hunting for ultra-light axions.  

Next, we consider the most relevant axion couplings for experimental sensitivity. In the low-energy effective Lagrangian for the axion, it is sometimes convenient to eliminate the axion coupling to the gluons via a field-dependent axial transformation of the SM quarks. After making such a rotation, the axion coupling to the photons is given by~\cite{GrillidiCortona:2015jxo}, 
\begin{align}\label{eq:axion_coupling_photon}
&\mathcal{L}\supset \underbrace{ \frac{\alpha_\mathrm{em}}{2\pi f_a}\left(\frac{\mathcal{\hat E}}{\mathcal{\hat N}}-1.92  \right)}_{\equiv g_{a\gamma\gamma}} \frac{a}{4} F \widetilde F, 
\end{align}
where the model-dependent quantity, apart from $f_a$ (see Eq.~\eqref{decayconstant}), in our case, is given by $\mathcal{\hat E}/\mathcal{\hat N}=8/3$. In fact, the dark matter experiment ABRACADABRA~\cite{Kahn:2016aff} has a great potential to look for an axion dark matter in the mass range of interest. As shown in Fig.~\ref{fig:DM}, a major part of the parameter space of our theory will be probed by this dark matter direct detection experiment. Fig.~\ref{fig:DM} is obtained by varying the model parameters while imposing all relevant constraints. The details of our numerical procedure are relegated to Sec.~\ref{sec:numerical_results}.  

Another axion dark matter experiment, the DMRadio-GUT~\cite{Domcke:2022rgu,DMRadio}, will also be sensitive in detecting axions with GUT scale decay constant $f_a(\sim 10^{16}$\,GeV). DMRadio-GUT will be far more sensitive compared to its previous two phases, i.e., DMRadio-50L and DMRadio-m$^3$, since it will have a factor
of three enhancement in the field and a factor of ten enhancement in volume relative to DMRadio-m$^3$.  The projected $3\,\sigma$ sensitivity of DMRadio-GUT  is also presented in Fig.~\ref{fig:DM} by a green shaded region, which will probe a significant portion of the parameter space. 
Yet another proposal  utilizing an optomechanical cavity~\cite{Murgui:2022zvy} filled with superfluid helium is shown to be highly promising in detecting  ultra-light axion dark matter. This proposed experimental method, with a cavity size of order $\mathcal{O}(10\,\mathrm{m})$ is expected to be sensitive to axion-photon couplings for axions with the GUT scale size decay constant. In Fig.~\ref{fig:DM}, the corresponding theoretical reach is shown with solid black lines. The ABRACADABRA experiment will be sensitive to axion masses as low as $m_a\sim 2$\,neV, whereas the sensitivity of DMRadio-GUT and optomechanical cavity is about $m_a\sim 0.4$\,neV and $m_a\sim 0.1$\,neV, respectively. A combination of all these axion dark matter experiments will eventually probe the entire parameter space of the proposed model.   

\begin{figure}[t!]
\centering
\includegraphics[scale=0.9]{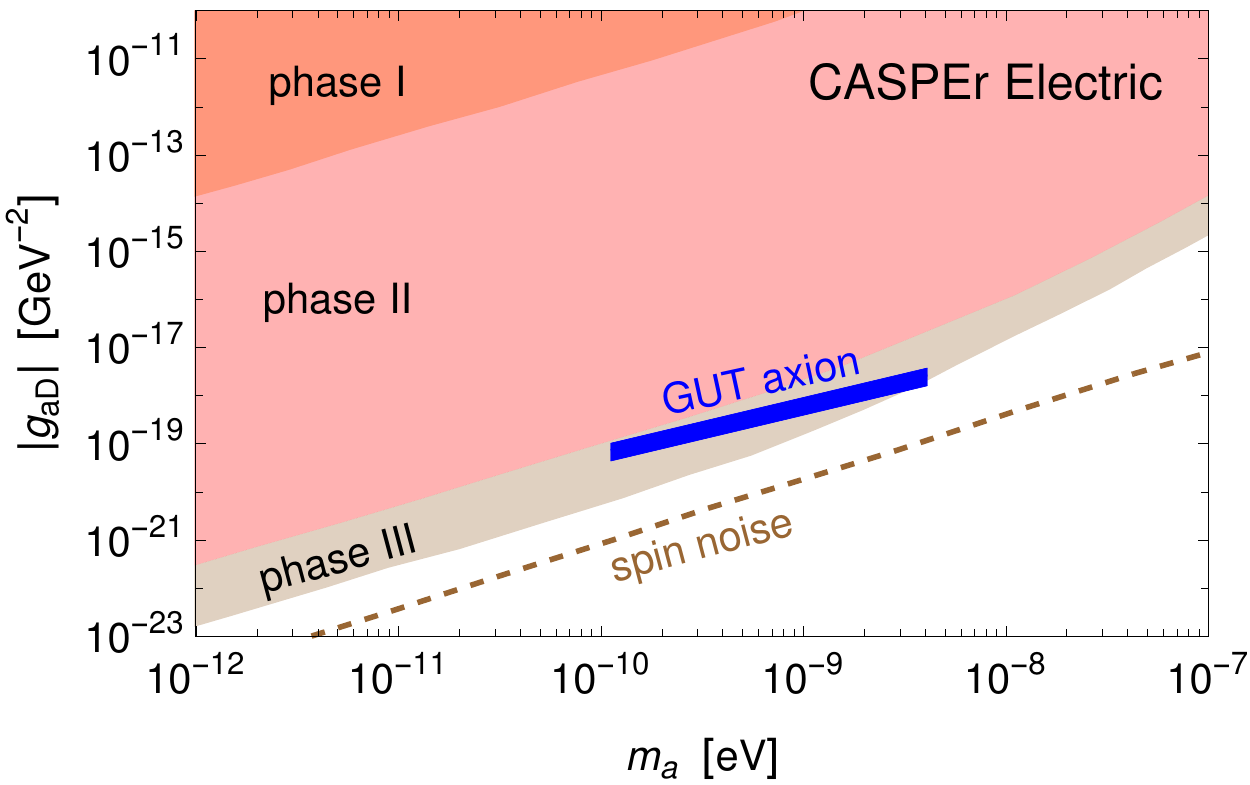}
\caption{  Axion coupling to the nucleon EDM operator  as a function of the axion mass.  The blue band  (that lies on the QCD line) corresponds to the prediction of our model; see text for details.  The shaded regions show the sensitivity projections of CASPEr Electric~\cite{Budker:2013hfa, JacksonKimball:2017elr} in its various phases.   Moreover, the ultimate sensitivity limit is given by the nuclear spin noise.    }\label{fig:EDM}
\end{figure}
Intriguingly, ultra-light axion dark matter can also be efficiently searched for via oscillating nucleon electric dipole moments (EDM). As already stated, the QCD axion solves the strong CP problem by promoting the $\theta$ parameter into the dynamical axion field.  Consequently, the effective $\theta$ angle gives rise to an EDM for nucleons sourced by the axion. Owing to the dynamical nature of the axion, this EDM will change in time, giving rise to unique signals. In the effective Lagrangian, the coupling of the axion to nucleon $n$ takes the following form,
\begin{align}
&\mathcal{L}\supset -\frac{i}{2}g_{aD}\, a\;\overline \psi_n \sigma_{\mu\nu}\gamma_5\psi_n F^{\mu\nu}\;.
\end{align}
The nucleon electric dipole moment generated through the above operator is given by $d_n=g_{aD}a$. The classical field that describes the axion field can be written as $a=a_0 \cos(m_a t)$. The amplitude, $a_0$, is determined from the local dark matter density, namely, $\rho_\mathrm{DM}=\frac{1}{2}m_a^2a^2_0$, which assumes  the axion comprises $100\%$ of dark matter within our setup. The nucleon electric dipole moment is then determined by the dark matter energy density, $d_n= \sqrt{2} g_{aD} \sqrt{\rho_\mathrm{DM}} \cos(m_a t)/m_a$. Moreover, the nucleon electric dipole moment can also be expressed in terms of the axion decay constant. In terms of our model parameters, it can be re-written in the following form~\cite{Graham:2013gfa}: 
\begin{align}\label{eq:nucleon_EDM}
    d_n \approx a\; \underbrace{ \frac{2.4\times 10^{-16}}{f_a}\,e\cdot \textrm{cm}}_{g_{aD}},
\end{align}
with roughly a 40\% uncertainty~\cite{Pospelov:1999ha}, where the decay constant is given in Eq.~\eqref{decayconstant}. (See also Refs.~\cite{Crewther:1979pi,Hisano:2012sc,Yoon:2017tag}.)
The corresponding coupling as a function of the axion mass is shown in Fig.~\ref{fig:EDM}. As can be seen from this figure, excitingly, the CASPEr Electric~\cite{Budker:2013hfa, JacksonKimball:2017elr} experiment alone will probe almost the entire parameter space of our model. The width of the band
corresponds to the calculation uncertainty as mentioned before. Fig.~\ref{fig:EDM} is also obtained by varying model parameters after one imposes all the relevant constraints. The exact details will be discussed later in the text. 

Since the axion is ultra-light in our setup, it can constitute the entirety of the dark matter. It is important to point out that  the breaking of the GUT symmetry to that of the SM gauge group $SU(5)\times U(1)_\mathrm{PQ} \to SU(3) \times SU(2) \times U(1)$ leads to an overproduction of super-heavy monopoles that must be inflated away. As discussed above, spontaneous breaking of the PQ symmetry leads to $N_\mathrm{DW}$ distinct degenerate vacua, giving rise to a domain-wall problem, which also requires dilution to be consistent with cosmology. Both of these problems, along with the horizon and flatness problems, can be elegantly solved via inflation taking place after the GUT symmetry breaking.  We, however, do not specify the details of the inflationary dynamics, which is beyond the scope of this work. The amount of axion dark matter produced then depends on whether the PQ symmetry is restored or not after inflation. We assume that the $U(1)_\mathrm{PQ}$ remains broken during inflation and is never restored afterwards. In such a scenario, the relic abundance of the axion dark matter is given by~\cite{Ballesteros:2016xej}
\begin{align}
\Omega h^2\sim   0.12\left( \frac{5\,\mathrm{neV}}{m_a} \right)^{1.17}
\left( \frac{\theta_i}{1.53\times 10^{-2}} \right)^2,
\end{align}
which shows that the initial value of $\theta_i=a_i/f_a$, where $a_i$ is the value of the axion field, needs to be somewhat smaller than unity to be consistent with the observed dark matter relic abundance $\Omega h^2\sim   0.12\pm 0.001$~\cite{Planck:2018vyg}.  Thus, for $\theta_i\sim 10^{-2}$, the axion is composed of all the dark matter.

\section{Unification, axion mass and proton decay}
\label{sec:numerical_results}

In our model, the axion decay width $f_a$ is connected to the GUT scale $M_\textrm{GUT}$ due to the fact that $24_H$ simultaneously breaks the $SU(5)$ and $U(1)_\textrm{PQ}$ symmetries. This, in particular, directly relates the axion mass $m_a$ to the GUT scale $M_\textrm{GUT}$ via Eq.~\eqref{eq:axion_mass}. Moreover, since the partial proton lifetimes are proportional to the fourth power of the GUT scale, our model can be simultaneously probed with axion dark matter and proton decay experiments.

\subsection{Unification}\label{sec:unification}
\begin{figure}
    \centering
    \includegraphics[width=0.95\textwidth]{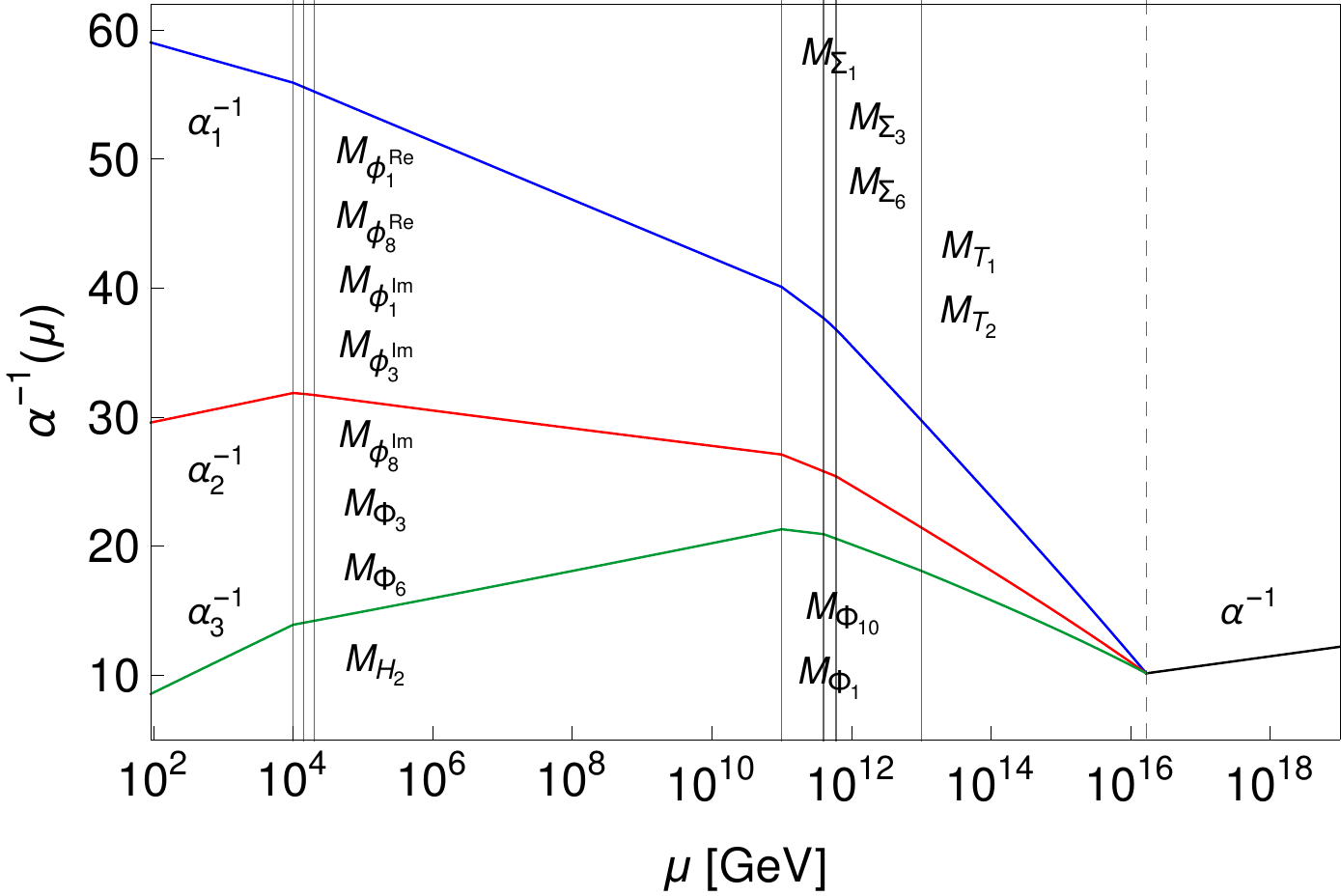}
    \caption{Example for the choice of the intermediate-scale particle masses giving gauge coupling unification.}
    \label{fig:gcu_example}
\end{figure}
The renormalization group equations (RGEs) for the gauge couplings can, at the 2-loop level, be written as \cite{Machacek:1983tz}
\begin{align}
\label{eq:2loop_gauge_running}
    \mu\frac{d\alpha^{-1}}{d\mu}=
    &-\frac{1}{2\pi}\left(b_i^{\textrm{SM}}+\sum_J b_i^J\mathcal{H}(\mu-M_J)\right)
    \nonumber\\
    &-\frac{1}{8\pi^2}\left(\sum_J\left(b_{ij}^{\textrm{SM}}+b_{ij}^J\mathcal{H}(\mu-M_J)\right)\alpha_j^{-1}+\beta_i^Y\right).
\end{align}
Here, $b_i^\textrm{SM}$ ($b_{ij}^\textrm{SM}$) are the SM 1-loop (2-loop) gauge coefficients, while $b_i^J$ ($b_{ij}^J$) are the 1-loop (2-loop) gauge coefficients of the multiplets $J$ with intermediate-scale masses $M_J$, i.e., $M_Z<M_J<M_\textrm{GUT}$. These coefficients are listed in Appendix~\ref{app:RGE}. Moreover, $\beta_i^Y$ are the Yukawa contributions and $\mathcal{H}$ is the Heaviside step function defined as
\begin{align}\label{eq-Heaviside step function}
    \mathcal{H}(m)=\left\{
\begin{array}{lll}
1, && m>0 \\
0, && \, m\leq 0 \\
\end{array}.
\right. 
\end{align} 
Note that we neglect the effect of the Yukawa couplings $Y^a,Y^b,$ and $Y^c$ on the running of the gauge couplings. 

In order to investigate the viable part of the parameter space giving gauge coupling unification, we freely vary the masses of the fields $\phi_1^\textrm{Re}$, $\phi_1^\textrm{Im}$, $\phi_3^\textrm{Im}$, $\phi_8^\textrm{Re}$, $\phi_8^\textrm{Im}$, $\Sigma_1$, $\Sigma_3$, $\Sigma_6$, $\Phi_1$, $\Phi_3$, $\Phi_6$, $\Phi_{10}$, and $H_2$ between the TeV and the GUT scale, while taking into account the mass spectrum constraints presented in Sec.~\ref{sec:model}. Some of these states remain light to achieve high scale unification to be compatible with proton decay bounds. To get an understanding of how these states can be light, let us consider the masses of $\phi_1^\textrm{Re}$ and $\phi_8^\textrm{Im}$, which are given by Eq.~\eqref{MASS1} and Eq.~\eqref{MASS2}, respectively. It can be easily seen that by adjusting the relevant quartic couplings, they can, in principle, live in the low energies (similar arguments are applicable for the rest of the states that reside somewhat below the GUT scale; for details, see Sec.~\ref{sec:model}). Such adjustments, however, introduce additional fine-turning problems on top of the usual doublet-triplet splitting problem, which we accept.  We also ensure that the scalar leptoquark mediated proton decay is sufficiently suppressed by varying the masses of $T_1$ and $T_2$ between $3\times 10^{11}$\,GeV and the GUT scale. The numerical fit is performed by running the gauge couplings at the 2-loop level from the GUT scale to the $Z$ mass scale at which a $\chi^2$-function that we define later in detail is minimized. We use the low-scale values $g_1=0.461425^{+0.000044}_{-0.000043}$, $g_2=0.65184^{+0.00018}_{-0.00017}$, and $g_3=1.2143^{+0.0035}_{-0.0036}$~\cite{Antusch:2013jca} as our input, where $g_i=\sqrt{4\pi\alpha_i}$. To demonstrate that within our setup, the gauge couplings can indeed unify, Fig.~\ref{fig:gcu_example} shows one possible particle mass spectrum giving exact gauge coupling unification that is in agreement with the current proton decay constraints and that yields correct neutrino mass scale via Eq.~\eqref{eq:massnu}.

\subsection{Proton decay}\label{sec:proton_decay}
The formulae for the proton decay widths of various decay channels can be found in Refs.~\cite{Claudson:1981gh,JLQCD:1999dld}. For example, the decay width for the proton decay channel having a pion and a charged lepton in the final state is given by\footnote{The Mathematica package \texttt{ProtonDecay}~\cite{Antusch:2020ztu} can be used to compute the decay widths of various nucleon decay channels.}
\begin{align}\label{eq:proton_decay}
    \Gamma(p\rightarrow \pi^0e_\alpha^+)&=\frac{ m_p\pi}{2}\left(1-\frac{m_\pi^2}{m_p^2}\right)^2A_L^2\frac{\alpha_\textrm{GUT}^2}{M_\textrm{GUT}^4}\\\nonumber
    &\times\left(A_{SL}^2|c(e_\alpha^c,d)\langle\pi^0|(ud)_Lu_L|p\rangle|^2+A_{SR}^2|c(e_\alpha,d^c)\langle\pi^0|(ud)_Ru_L|p\rangle|^2\right).
\end{align}
Here, $m_p=0.9393$\,GeV and $m_\pi=0.134$\,GeV denote the proton and pion masses, respectively, while $\alpha =1,2$ with $e_1^+ \equiv e^+$ and $e_2^+ \equiv \mu^+$. The leading log renormalization of the dimension six operators is encoded via $A_L= 1.2$\ \cite{Nihei:1994tx} and $A_{SL(R)}$, where\footnote{If the denominator of the exponent vanishes for some factor, i.e., the 1-loop running of a specific gauge coupling is constant within a certain interval, the respective factor in Eq.\ \eqref{eq:ASL(R)} is replaced with $\exp[\gamma_{L(R)i}\alpha(M_{I+1})]/(2\pi)$.}
\begin{align}\label{eq:ASL(R)}
    A_{SL(R)}=\prod_{i=1,2,3}\prod_{I}^{M_Z\leq M_I\leq M_\textrm{GUT}}\left(\frac{\alpha_i(M_{I+1})}{\alpha_i(M_I)}\right)^{\frac{\gamma_{L(R)i}}{b_i^\textrm{SM}+\sum_J^{M_Z\leq M_J\leq M_I}b_i^J}},
\end{align}
with $\gamma_{L(R)i}=\left(23(11)/20,9/4,2\right)$\ \cite{Wilczek:1979hc,Buras:1977yy,Ellis:1979hy}.
Moreover, we take the hadron matrix elements, such as, for example, $\langle\pi^0|(ud)_Lu_L|p\rangle=0.134(5)(16)$\,GeV$^2$ and $\langle\pi^0|(ud)_Ru_L|p\rangle=-0.131(4)(13)$\,GeV$^2$, from Refs.~\cite{Aoki:2017puj,Yoo:2021gql}. Finally, the c-coefficients of Eq.~\eqref{eq:proton_decay} read~\cite{DeRujula:1980qc,FileviezPerez:2004hn,Nath:2006ut}
\begin{align}
    &c(e_\alpha^c,d_\beta)=(U_R^\dagger U_L^\ast)_{11}(E_R^\dagger D_L^\ast)_{\alpha\beta}+(E_R^\dagger U_L^\ast)_{\alpha 1}(U_R^\dagger D_L^\ast)_{1\beta}\ ,\\
    &c(e_\alpha,d^c_\beta)=(U_R^\dagger U_L^\ast)_{11}(E_L^\dagger D_R^\ast )_{\alpha\beta}\ ,\\
    &c(\nu_l,d_\alpha,d^c_\beta)=(U_R^\dagger D_L^\ast)_{1\alpha}(D_R^\dagger N)_{\beta l}\ ,
\end{align}
where the unitary matrices $U_{L/R}$, $E_{L/R}$, $D_{L/R}$, and $N$ diagonalize the SM fermion mass matrices through the following transformations
\begin{align}\label{eq:mixing_matrices}
    &M_u=U_L M_u^\textrm{diag}U_R^\dagger,    && M_d=D_L M_d^\textrm{diag}D_R^\dagger,   \nonumber\\
    &M_e=E_L M_e^\textrm{diag}E_R^\dagger,    && M_\nu=N M_\nu^\textrm{diag}N^T.
\end{align}

The current experimental constraints and future sensitivities for the various partial lifetimes that we use in our numerical analysis are presented in Table~\ref{tab:nucleon_decay}. For a recent review on the subject, see Ref.~\cite{Dev:2022jbf}.

\begin{table}[t!]
\centering
\begin{tabular}{|c|c|c|}\hline
decay channel  & current bound $\tau_p$ [yrs] & future sensitivity $\tau_p$ [yrs] 
\\\hline\hline
$p\rightarrow \pi^0\,e^+$  &  $2.4\cdot 10^{34}$ \cite{Super-Kamiokande:2020wjk} &  $7.8\cdot 10^{34}$ \cite{Hyper-Kamiokande:2018ofw}   \\\hline
$p\rightarrow \pi^0\,\mu^+$  &  $1.6\cdot 10^{34}$ \cite{Super-Kamiokande:2020wjk} &  $7.7\cdot 10^{34}$ \cite{Hyper-Kamiokande:2018ofw}    \\\hline
$p\rightarrow \eta^0\,e^+$  &  $1.0\cdot 10^{34}$ \cite{Super-Kamiokande:2017gev} &  $4.3\cdot 10^{34}$ \cite{Hyper-Kamiokande:2018ofw}  \\\hline
$p\rightarrow \eta^0\,\mu^+$  &  $4.7\cdot 10^{33}$ \cite{Super-Kamiokande:2017gev} &  $4.9\cdot 10^{34}$ \cite{Hyper-Kamiokande:2018ofw}   \\\hline
$p\rightarrow K^0\,e^+$  &  $1.1\cdot 10^{33}$ \cite{Brock:2012ogj} &  -  \\ \hline
$p\rightarrow K^0\,\mu^+$  &  $3.6\cdot 10^{33}$ \cite{Super-Kamiokande:2022egr} &  -  \\\hline
$p\rightarrow \pi^+\,\overline{\nu}$  &  $3.9\cdot 10^{32}$ \cite{Super-Kamiokande:2013rwg} &  -  \\\hline
$p\rightarrow K^+\,\overline{\nu}$  &  $6.6\cdot 10^{33}$ \cite{Takhistov:2016eqm} &  $3.2\cdot 10^{34}$ \cite{Hyper-Kamiokande:2018ofw} \\
\hline
\end{tabular}
\caption{Present experimental bounds on the partial lifetimes $\tau_p$ as well as future sensitivities for 10 years of runtime, both at 90\% confidence level. }\label{tab:nucleon_decay}
\end{table}

\subsection{Numerical procedure}\label{sec:numerical_procedure}
We start our numerical analysis by constructing matrices $M_u$, $M_e$, $Y^a$, $Y^b$, and $Y^c$ at the GUT scale, as described in the next few paragraphs.  

Since the up-type quark mass matrix $M_u$ is approximately symmetric, we have that $U_R=U_L^\ast$. This allows us to express $M_u$ as 
\begin{align}
M_u=U_L \textrm{diag}(m_u,m_c,m_t) U_L^T.
\end{align}
We furthermore parametrize the up-type quark mixing matrix $U_L$ in terms of the down-type quark mixing matrix $D_L$, the Cabibbo-Kobayashi-Maskawa (CKM) matrix $V_\textrm{CKM}$, and five GUT phases $\beta^u_1$, $\beta^u_2$, $\eta^u_1$, $\eta^u_2$, and $\eta^u_3$, as
\begin{align}
    U_L=D_L\textrm{diag}(e^{i\beta^u_1},e^{i\beta^u_2},1)V_\textrm{CKM}^T\textrm{diag}(e^{i\eta^u_1},e^{i\eta^u_2},e^{i\eta^u_3})\ .
\end{align}
In our analysis, we set $\eta^u_1=\eta^u_2=\eta^u_3=0$ since these three phases do not affect the proton decay predictions at all.

We set $E_L=E_R=\mathds{1}$ since $M_e$ is diagonal and real. This also means that we can simply construct $M_e$ via an equality that reads  
\begin{align}
    M_e=\textrm{diag}(m_e,m_\mu,m_\tau). 
\end{align}

$Y^a$ and $Y^b$ are constructed via Eq.~\eqref{eq:perturbativity} using the neutrino mixing matrix $N=\textrm{diag}(e^{i\eta^\nu_1},e^{i\eta^\nu_2},e^{i\eta^\nu_3})V_\textrm{PMNS}^\ast$ as an input. Note that $V_\textrm{PMNS}$ contains the CP violating phase $\delta^\nu$ as well as the Majorana phase $\beta^\nu$. We furthermore take $Y^c$ to be a general complex $1\times 3$ matrix through 
\begin{align}
    Y^c=(y^c_1e^{i\eta^c_1} \quad y^c_2e^{i\eta^c_2} \quad y^c_3e^{i\eta^c_3})^T.
\end{align}
Once the parameter dependence of $M_u$, $M_e$, $Y^a$, $Y^b$, and $Y^c$ is properly accounted for, as described above, we can also construct $M_d$ and $M_\nu$ that are given by Eqs.~\eqref{eq:massd} and \eqref{eq:massnu}, respectively. We treat $\lambda$ in $M_\nu$ as a free parameter while the two Higgs VEVs that enter $M_d$ and $M_\nu$ are given by $v_{\Lambda_1}=v_{\Lambda_2}=174/\sqrt{2}$\,GeV due to the constraint that $\tan \beta$ of Eq.~\eqref{eq:tanbeta} is equal to one.

In summary, the free parameters for our numerical analysis are the unification scale $M_\textrm{GUT}$ and the corresponding gauge coupling $\alpha_\textrm{GUT}$, the masses of the fields\footnote{Note that the masses of the fields $\phi_1^\textrm{Re}$, $\phi_1^\textrm{Im}$, $\Sigma_3$, $\Sigma_6$, $\Phi_{10}$ are obtained via the mass relations discussed in Sec.~\ref{sec:model}.}  $\phi_3^\textrm{Im}$, $\phi_8^\textrm{Re}$, $\phi_8^\textrm{Im}$, $\Sigma_1$,  $\Phi_1$, $\Phi_3$, $\Phi_6$, $T_1$, $T_2$, and $H_2$, the phases $\beta^u_{1,2}$,  $\delta^\nu$, $\beta^\nu$, $\eta^\nu_{1,2,3}$, the Yukawa parameters $y^c_{1,2,3}$,  $\eta^c_{1,2,3}$, the quartic Higgs coupling $\lambda$, and the scaling parameter $\xi$. These 24 parameters are fitted to the experimental observables that are the SM gauge couplings $g_1$, $g_2$, and $g_3$, and the down-type quark masses $m_d$, $m_s$, and $m_b$, while requiring that the current proton decay constraints, as given in Table~\ref{tab:nucleon_decay}, are satisfied. Note that the charged lepton masses, the up-type quark masses, the neutrino mass squared differences, the CKM mixing parameters, and the known PMNS mixing parameters are all automatically accounted for. 

Since there are more parameters than observables, proton decay cannot be predicted sharply in all decay channels as we will discuss in the next section. But, due to the fact that the neutrino mass matrix is connected to the mismatch between the charged lepton and down-type quark mass matrices, our model predicts the PMNS parameters $\delta^\nu$ and $\beta^\nu$ to be in relatively narrow intervals.

The gauge couplings are fitted to their low-energy scale values~\cite{Antusch:2013jca} after the 2-loop level running from the high scale to the low scale is performed. To simplify the analysis, we do not run the Yukawa parameters from low scale to the GUT scale using RGEs, and  the down-type quark and neutrino masses are directly fitted at the high scale using the high scale values provided in Ref.~\cite{Babu:2016bmy}. The $\chi^2$-function is obtained comparing the theoretical prediction $p_i$ with the experimental central value $e_i$, normalized with the corresponding experimental standard deviation $\sigma_i$ of the $i$-th observable via
\begin{align}
    \chi^2=\sum_i\left(\frac{p_i-e_i}{\sigma_i}\right)^2.
\end{align}
To minimize the $\chi^2$-function we apply a differential evolution algorithm. This minimization yields a satisfactory benchmark point and thus proves the viability of our model. Then, starting from this benchmark point, a Markov-chain-Monte-Carlo (MCMC) analysis with a flat prior distribution, involving a Metropolis-Hasting algorithm, is performed giving us a total of $6\times10^6$ datapoints. Finally, we use these points to calculate the highest posterior density (HPD) regions of various quantities. 

For the numerical analysis, all parameters are freely varied in such a way that the perturbativity of all Yukawa and Higgs couplings is satisfied. In particular, the absolute values of all entries in $Y^a$, $Y^b$, and $Y^c$ as well as the absolute value of $\lambda$ are all required to be less than or equal to 1. To this end, the scaling parameter $\xi$ ensures that the full parameter space is covered with the chosen parametrization of the matrices $Y^a$ and $Y^b$. Furthermore, although we fix some model parameters during the fitting/minimization procedure by directly plugging in experimental central values of some observables, we still vary these parameters in the subsequent MCMC analysis.

\begin{figure}[th!]
\centering
\includegraphics[width=0.5\textwidth]{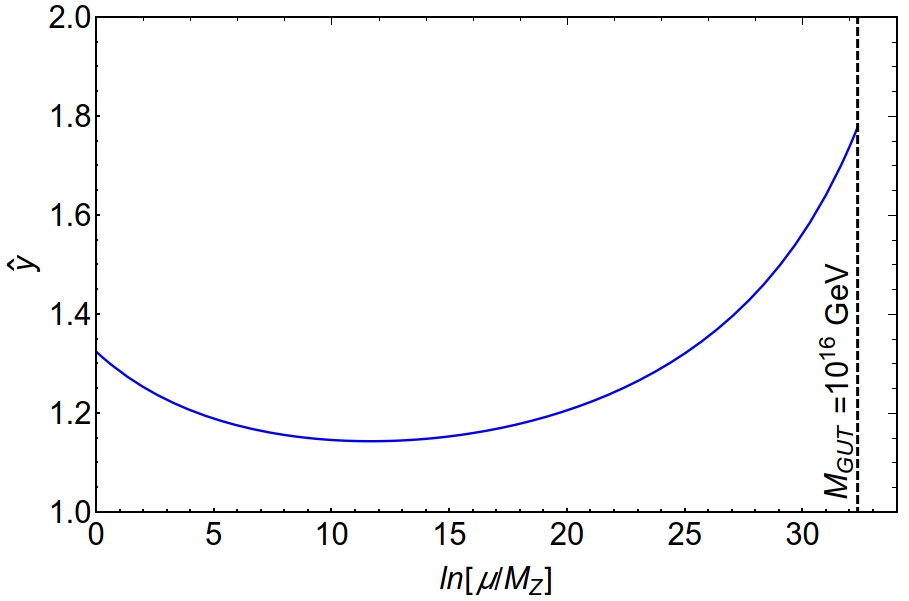}
\caption{RGE running of the Yukawa coupling that is relevant for to top quark mass generation. See text for details.}\label{fig:top}
\end{figure}

It is not necessary to fit the up-type quark masses in our numerical analysis, as already mentioned above. We, however,  briefly address perturbativity of the top quark Yukawa coupling. The Yukawa couplings, once the GUT symmetry is broken down to the SM gauge group, need to be matched with the Yukawa couplings of the effective theory, which in our case resembles the type-II 2HDM (two Higgs doublet model). Thus, in the initial basis, the up-type quark sector interacts with one Higgs doublet whereas the down-type quark and charged lepton sectors interact with the other Higgs doublet. Consequently, the SM top quark Yukawa coupling reads $y^\mathrm{SM}_t = \cos\beta \hat y$, where $\hat y$ is the coupling we are interested in while $\cos\beta=1/\sqrt{2}$. To investigate the perturbativity of $\hat y$ we need to study the following RGEs~\cite{Branco:2011iw}
\begin{align}
16\pi^2\mu \frac{d}{d\mu} \hat y=\Bigg\{ \underbrace{ \left( -8g^2_3-\frac{9}{4}g^2_2 - \frac{17}{12}g^2_1 \right) }_\mathrm{negative}  + \underbrace{ \frac{9}{2} \hat y^2 }_\mathrm{positive}\Bigg\} \hat y,  \label{eq:top}
\end{align}
and 
\begin{align}
&16\pi^2\mu \frac{d}{d\mu} g_k =\underbrace{ c_k }_{(c_3, c_2, c_1)_\mathrm{2HDM}=(-7,-3,7)} g^3_k.
\end{align}
Here, for simplicity, we only consider an effective theory of 2HDM. We evolve these RGEs from the $M_Z$ scale to the GUT scale, which we choose to be $10^{16}$\,GeV. The initial value of the top quark Yukawa is extracted from the value of the top quark mass in the $\overline{\mathrm{MS}}$ scheme, i.e., $\overline m_t (\overline m_t)=163$ GeV~\cite{Hoang:2020iah}, providing us with $\hat y=\overline m_t/(v\cos\beta)$, where $v=174.104$\,GeV. (Recall that $\overline m_t$ is the scale-dependent mass and not the physical mass of the top-quark~\cite{Hoang:2020iah}.) Our result for the running of $\hat y$ is presented in Fig.~\ref{fig:top}. Clearly, the coupling $\hat y$ remains perturbative up to the GUT scale due to an interplay between the gauge and Yukawa coupling contributions of Eq.~\eqref{eq:top}.  

To generate Fig.~\ref{fig:top}, we solely consider an effective 2HDM scenario. Our scenario, however, is more complex since several scalar multiplets live significantly below the GUT scale. These fields affect the running of the gauge couplings, as can be seen in Fig.~\ref{fig:gcu_example}, and make the couplings substantially larger, at higher energy scales, when compared to the 2HDM case. This simply means that the negative contributions in Eq.~\eqref{eq:top} are even more important than in the 2HDM case and can thus easily result in smaller coupling $\hat y$ at the high energy scale, if compared to naive expectation. This effect has already been pointed out in Ref.~\cite{Dorsner:2021qwg}, where the RGE running of the charged fermion Yukawa couplings has been implemented.

\subsection{Results}
In this section, we present the outcome of our numerical study. We are interested in the full axion mass range, the predictions for partial proton decay lifetimes, and the viable range of the Dirac CP and Majorana phases of the PMNS matrix.

The axion mass $m_a$ is connected to the GUT scale $M_\textrm{GUT}$ and gauge coupling $\alpha_\textrm{GUT}$ via Eq.~\eqref{eq:axion_mass}. We can therefore obtain the predicted range of the axion mass by maximizing and minimizing Eq.~\eqref{eq:axion_mass}. We demand viable gauge coupling unification and correct neutrino mass scale while making sure that none of the current proton decay constraints are violated. We find $m_a\in[0.1,4.7]$\,neV which we present in Figs.~\ref{fig:DM} and \ref{fig:EDM}. As discussed in more detail in Sec.~\ref{sec:Peccei-Quinn}, this already demonstrates that the full parameter space will be probed by two kinds of future axion DM experiments that are sensitive to either the axion to photon coupling or to the nucleon EDM. 

To start our numerical analysis, we find a viable benchmark point from a full $\chi^2$ fit. In particular, for the case of normal neutrino mass ordering, we obtain that 
\begin{align}\label{eq:Ya_Yb_Yc}
    &Y^{a}=\begin{pmatrix}
       -0.120 + i\,0.00943, & 0.513 + i\,0.200, & 0.898 
    \end{pmatrix},\\
    &Y^{b}=\begin{pmatrix}
        0.109 + i\,0.150, & 0.348 + i\,0.334, & 0.195 - i\,0.0211
    \end{pmatrix},\\        \label{eq:Yc}
    &Y^{c}=\begin{pmatrix}
        0.00115 +i\,0.00198 , & -0.0532+i\,0.0852, & -2.781-i\,0.743
    \end{pmatrix}\times 10^{-6},
\end{align}
for $M_\textrm{GUT}=10^{16.2}$\,GeV, $m_{H_2}=10^{3.77}$\,GeV, $M_{T_1}=M_{T_2}=10^{14.55}$\,GeV, $M_{\phi_1^\textrm{Re}}=10^{4.39}$\,GeV, $M_{\phi_1^\textrm{Im}}=10^{4.12}$\,GeV, $M_{\phi_3^\textrm{Im}}=10^{4.40}$\,GeV, $M_{\phi_8^\textrm{Re}}=10^{4.09}$\,GeV, $M_{\phi_8^\textrm{Im}}=10^{3.71}$\,GeV, $M_{\Sigma_1}=10^{13.41}$\,GeV, $M_{\Sigma_3}=10^{12.63}$\,GeV, $M_{\Sigma_3}=10^{13.24}$\,GeV, $M_{\Phi_1}=10^{11.63}$\,GeV, $M_{\Phi_3}=10^{5.28}$\,GeV, $M_{\Phi_6}=10^{4.18}$\,GeV, $M_{\Phi_{10}}=10^{11.63}$\,GeV, $\alpha_\textrm{GUT}^{-1}=15.62$, and $\lambda=1.00$.
If the proton decay pull is neglected, this choice of the input parameters gives $\chi^2$ below 0.01. This is thus a perfect fit for the gauge couplings as well as for the fermion masses and mixings.  It is to be pointed out that even though the benchmark point presented here corresponds to a scenario with $T_1$ and $T_2$ masses being two orders of magnitude smaller than the GUT scale, these fields can easily reside at the GUT scale without significantly affecting the value of $M_\mathrm{GUT}$.

The PMNS Dirac CP phase, for this benchmark point, is given by $\delta^\nu=-48.5^\circ$, whereas the PMNS Majorana phase is $\beta^\nu=-71.3^\circ$. We note that for the case of inverted neutrino mass ordering, no good fit-point can be obtained. This is due to the fact that the Yukawa matrix $Y^a$ is needed to generate both the viable neutrino masses and the correct mismatch between the charged lepton and down-type quark masses. In the case of inverted ordering, the first two entries in $Y^a$ would need to be somewhat larger than the third entry. This is, however, in conflict with the down-type quark mass fit that requires the first entry of $Y^a$ to be smaller than the second and third entries. Therefore, our model predicts that the neutrinos have normal mass ordering.

\begin{figure}[t!]
\centering
\includegraphics[width=0.7\textwidth]{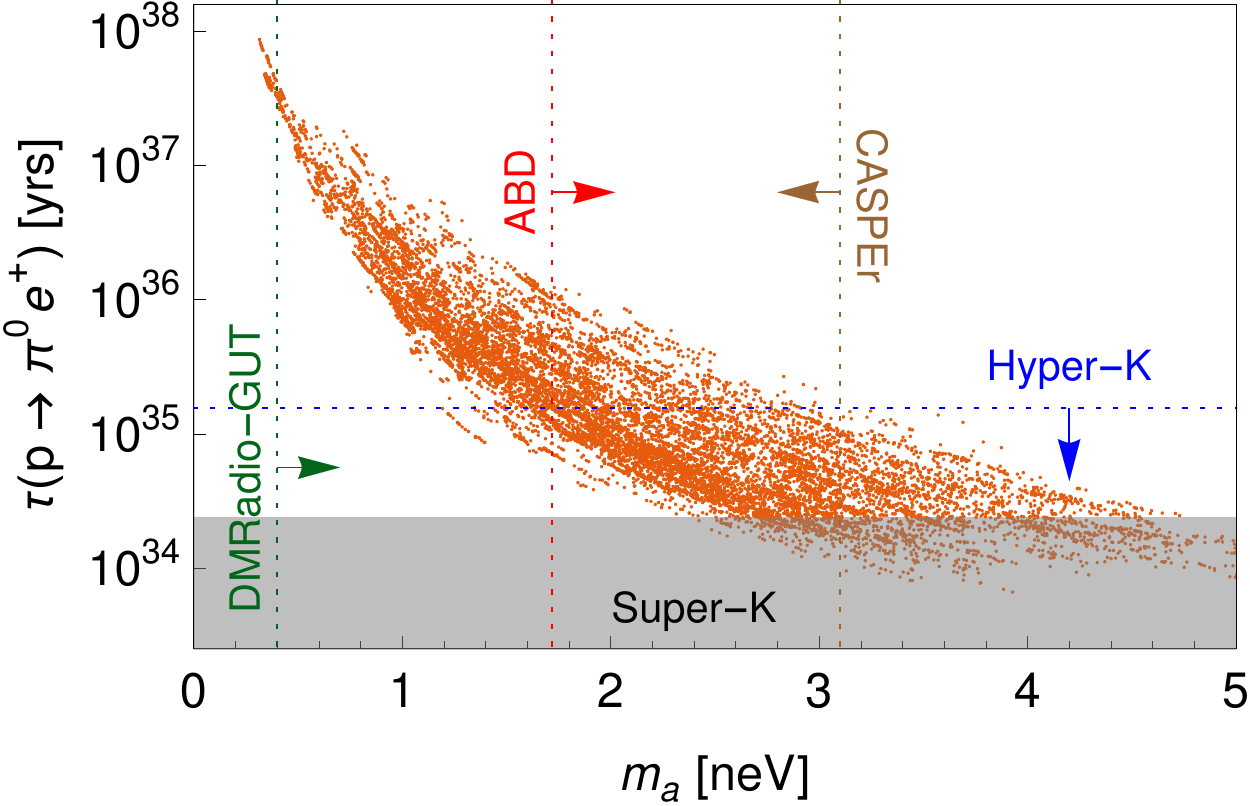}
\caption{The generated points from the MCMC analysis presented in the $m_a-\tau(p\rightarrow \pi^0e^+)$ plane. The current Super-Kamiokande bound is represented by a gray box, while the future Hyper-Kamiokande sensitivity is indicated by a blue dotted line. Moreover, the projected sensitivity of various axion DM experiments is also shown: ABRACADABRA (ABD) with a red dotted line, DMRadio-GUT with a green dotted line, CASPEr Electric with a brown dotted line. For details, see the main text. }\label{fig:ma_vs_pd}
\end{figure}

From the aforementioned benchmark point, we start an MCMC analysis with a flat prior. All obtained points are presented in Fig.~\ref{fig:ma_vs_pd} in a plane of axion mass vs.\ partial proton decay lifetime in the dominant decay channel $p\rightarrow \pi^0e^+$. We also present the future sensitivities of the DM experiments ABRACADABRA, DMRadio-GUT, and CASPEr Electric,  as discussed in Sec.~\ref{sec:Peccei-Quinn}, as well as the future sensitivity of the proton decay experiment Hyper-Kamiokande, as discussed in Sec.~\ref{sec:proton_decay}. Fig.~\ref{fig:ma_vs_pd} nicely visualizes how various parts of the model parameter space can be probed through the synergy between three different kinds of experiments testing ($i$) the axion to photon coupling, ($ii$) the nucleon EDM, and ($iii$) proton decay. For example, if the axion mass is observed to be above 3\,neV, proton decay via $p\rightarrow \pi^0e^+$ necessarily has to be seen by Hyper-Kamiokande if our model is realized in nature. Moreover, regardless of whether proton decay will be observed by Hyper-Kamiokande, the former two kinds of experiments will be able to cover the entire parameter space of our model.

\begin{figure}[p!]
    \centering
    \includegraphics[width=\textwidth]{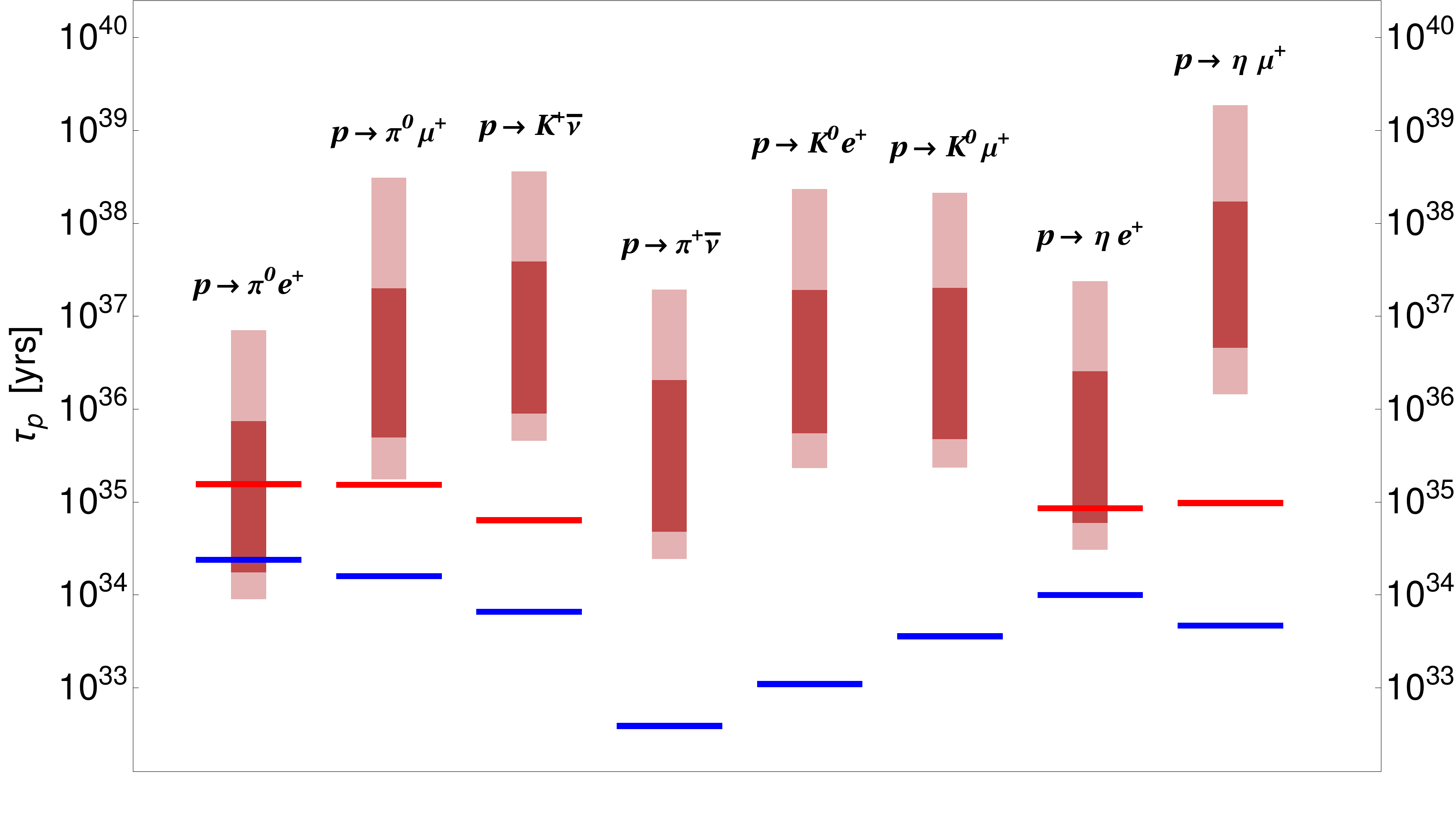}
    \caption{The predicted 1\,$\sigma$ (dark) and 2\,$\sigma$ (light) HPD intervals of the proton lifetime for various decay channels. The blue (red) line segments indicate the current (future) experimental bounds (sensitivities) at 90\% confidence level. Interestingly, a part of the predicted 1\,$\sigma$ region for both decay channels $p\rightarrow \pi^0e^+$ and $p\rightarrow \eta^0e^+$ lies within the reach of Hyper-Kamiokande.}\label{fig:pd_full} 
    \vspace{8mm}
    \includegraphics[width=\textwidth]{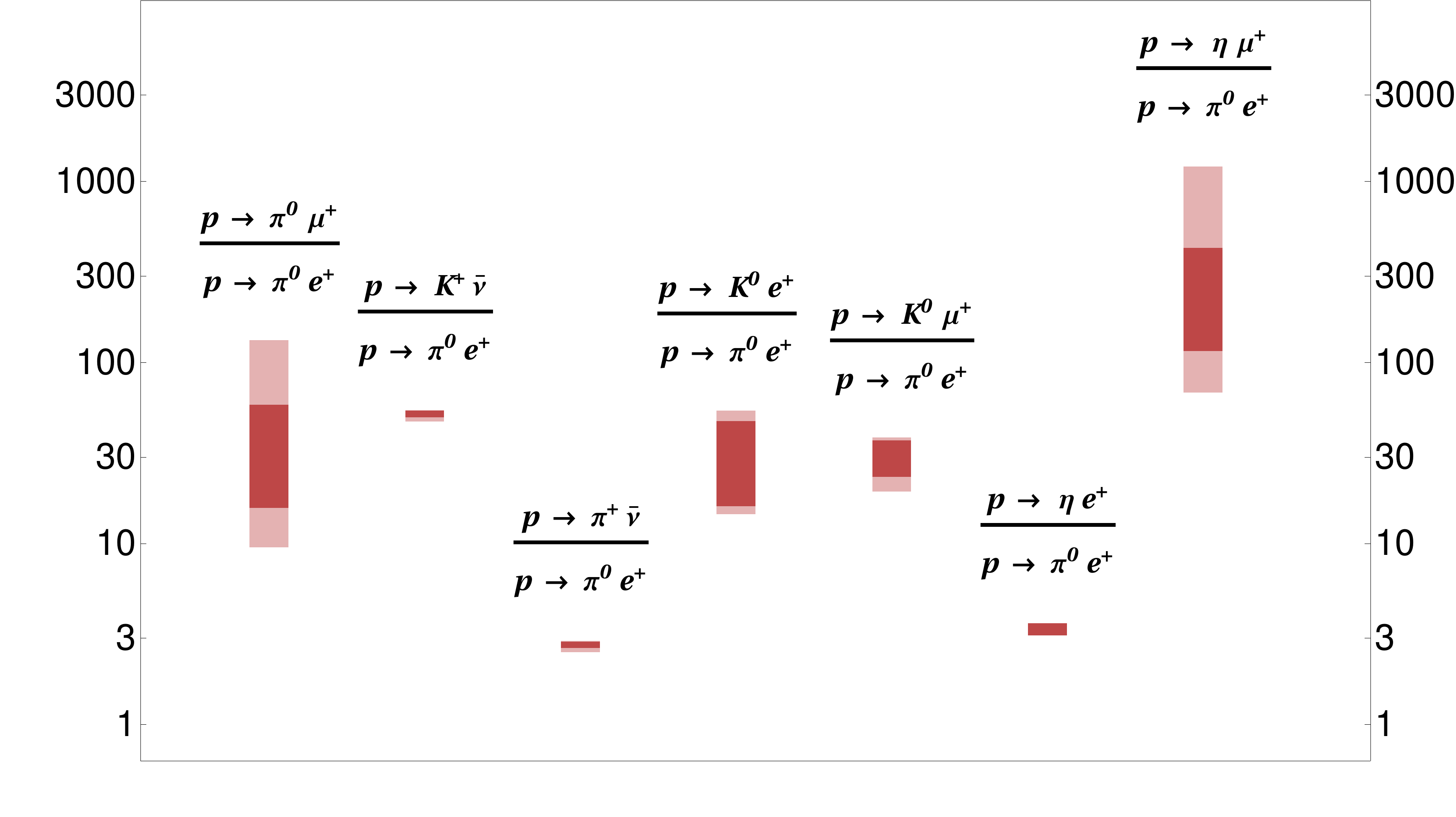}
    \caption{The 1\,$\sigma$ (dark) and 2\,$\sigma$ (light) HPD intervals of ratios of the proton lifetime of various decay channels. Interestingly, the ratio $\tau({p\rightarrow \eta^0e^+})/\tau({p\rightarrow \pi^0e^+})$ (which will partly be tested by Hyper-Kamiokande) is predicted very sharply.}
    \label{fig:pd_ratios}
\end{figure}

We are also interested in the proton decay predictions of all two-body decay channels within our model. First, we want to obtain the full allowed range for all partial proton lifetimes, which is for the decay channel $p\rightarrow \pi^0e^+$ already hinted in Fig.~\ref{fig:ma_vs_pd}. To this end, we vary all the parameters, including the intermediate-scale particle masses in the MCMC analysis. The 1\,$\sigma$ (dark) and 2\,$\sigma$ (light) HPD results of this analysis are shown in Fig.~\ref{fig:pd_full}. The blue line segments indicate the current experimental bounds, while the red line segments represent the future sensitivities. (See, for example, Table~\ref{tab:nucleon_decay}.) Fig.~\ref{fig:pd_full} shows that a part of the predicted 1\,$\sigma$ HPD interval for the two decay channels $p\rightarrow \pi^0e^+$ and $p\rightarrow \eta^0e^+$ will be tested by Hyper-Kamiokande. The large uncertainty in these partial lifetime predictions that are coming from the dependence on the fourth power of the GUT scale can be erased by considering ratios of specific decay channels.\footnote{For recent works analyzing ratios of partial proton decay lifetimes in models with predicted GUT scale quark-lepton Yukawa ratios, see Refs.~\cite{Antusch:2020ztu,Antusch:2021yqe}.} Fig.~\ref{fig:pd_ratios} shows the prediction of such ratios with the dominant decay channel $p\rightarrow \pi^0e^+$ in the denominator. Especially interesting is the prediction for the ratio $\tau({p\rightarrow \eta^0e^+})/\tau({p\rightarrow \pi^0e^+})$, since both $\tau({p\rightarrow \eta^0e^+})$ and $\tau({p\rightarrow \pi^0e^+})$ will be partly tested by Hyper-Kamiokande. This ratio is predicted very sharply. However, such a sharp prediction for this particular ratio is not only specific to our model but a more common feature of models in which gauge boson mediated proton decay is dominant and in which the contribution involving the c-coefficient $c(e^c,d)$ dominates the contribution with the c-coefficient $c(e,d^c)$. Nevertheless, these ratios of partial proton lifetimes provide an interesting additional opportunity to probe our model.

\begin{figure}[t!]
    \centering
     \includegraphics[width=0.95\textwidth]{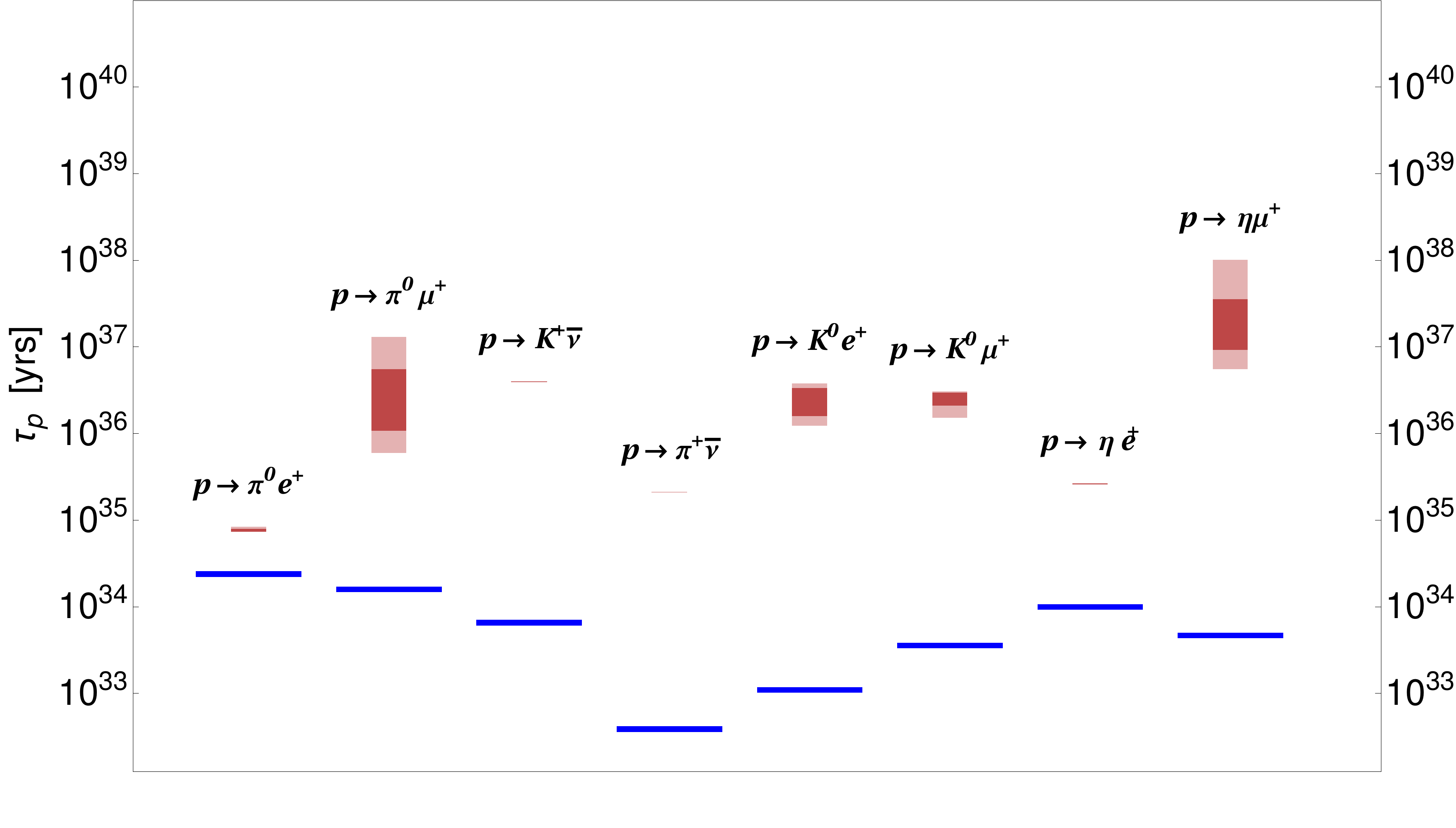}
    \caption{The 1\,$\sigma$ (dark) and 2\,$\sigma$ (light) HPD intervals of the proton lifetime for various decay channels for a benchmark scenario with   $M_\textrm{GUT}=10^{16.2}$\,GeV. The blue line segments represent the current experimental bounds at 90\% confidence level.    }
    \label{fig:pd_fixed_mgut}
\end{figure}

In order to understand the dependence of different decay channels on the flavor structure in the fermion mass matrices, we fix the mass scales to the same values as listed below Eq.~\eqref{eq:Yc} and only vary the parameters in the fermion mass matrices in the MCMC analysis, computing for each point the proton decay prediction for individual decay channels. We visualize the 1\,$\sigma$ (dark) and 2\,$\sigma$ (light) HPD results of this analysis in Fig.~\ref{fig:pd_fixed_mgut}, where the blue line segments indicate the current experimental bounds at 90\% confidence level presented in Table~\ref{tab:nucleon_decay}. Interestingly, the partial lifetimes, for some channels, are much more sharply predicted than for the others. The sharp prediction for the decay channels with an antineutrino in the final state is a generic feature for models with a (nearly) symmetric up-type quark mass matrix. On the other hand, the fact that the partial lifetime of the decay channel $p\rightarrow \pi^0e^+$ has such a sharp prediction is uncommon and represents a nice feature of our model, which also implies that this decay channel is predicted to be the dominant one.\footnote{Note, however, that proton decay mediated by the two scalar triplets $T_1$ and $T_2$ could enhance the decay channel $p\rightarrow K^+\overline{\nu}$.} 

The interesting result that some decay channels yield much sharper predictions than others can be understood by investigating the freedom in the mixing matrices that are defined in Eq.~\eqref{eq:mixing_matrices}. This is demonstrated in the following example, where we compare the predictions for the $p\rightarrow \pi^0e^+$ and $p\rightarrow \pi^0\mu^+$ lifetimes. The relevant c-coefficients of the two decay channels in question read
\begin{align}
    &c(e^c_\alpha,d)=(D_L^\ast)_{\alpha1}+(U_L^\ast)_{\alpha1}(U_L^TD_L^\ast)_{11},    \label{eq:c-coeff1}\\
    &c(e_\alpha,d^c)=(D_R^\ast)_{\alpha1}.     \label{eq:c-coeff2}
\end{align}
As it can be seen from Eq.~\eqref{eq:massd}, the left mixing of the down-type quark mass matrix $D_L$ strongly depends on the Yukawa matrix $Y^c$, while the right mixing $D_R$ dominantly depends on the Yukawa matrix $Y^a$. Since $Y^a$ has to be chosen in such a way that the correct PMNS parameters and neutrino masses are obtained, there cannot be a strong hierarchy between $Y^a$ entries. On the other hand, a strong hierarchy of the entries in $Y^c$ is required in order to produce the correct mismatch between the down-type quark and charged lepton masses. Therefore, $D_R$ appears to have a large mixing, whereas $D_L$ is for all points in the MCMC almost equal to the identity matrix. This, in particular, also implies that the CKM mixing is mostly coming from $U_L$. Hence, in the case of a positron in the final state, i.e., for $\alpha=1$ in Eqs.~\eqref{eq:c-coeff1} and \eqref{eq:c-coeff2}, the contribution coming from $c(e^c,d)$ dominates over the contribution coming from $c(e,d^c)$ in the decay width formula (see Eq.\ \eqref{eq:proton_decay}), since $|(U_L)_{11}|,|(D_L)_{11}|>|(D_R)_{11}|$. Contrarily, if an antimuon is in the final state ($\alpha=2$), the contribution involving $c(\mu,d^c)$ is dominant over the contribution from $c(\mu^c,d)$, since $|(D_R)_{21}|>|(U_L)_{21}|,|(D_L)_{21}|$. Now, varying over the full flavor freedom, since we always roughly have $|(U_L)_{11}|\approx|(V_\textrm{CKM})_{11}|,\,|(D_L)_{11}|\approx 1$, the dominating c-coefficient $c(e^c,d)$ only varies by an order 1 factor. This results in a very sharp prediction for the partial lifetime of the decay channel $p\rightarrow \pi^0e^+$. On the other hand, $|(D_R)_{21}|$ roughly varies within the interval $[0.1,1]$, resulting in a much less sharp prediction for the partial lifetime of the decay width $p\rightarrow \pi^0\mu^+$. 

Finally, from our MCMC results, we deduce the HPD intervals of the Dirac CP and Majorana phase of the PMNS matrix. At 1\,$\sigma$ we obtain $\delta^\nu\in[-22.6^\circ,34.4^\circ]$ and $\beta^\nu\in[-124.1^\circ,-71.4^\circ]$, while our 2\,$\sigma$ HPD results are $\delta^\nu\in[-50.7^\circ, 55.6^\circ]$ and $\beta^\nu\in[-132.2^\circ,-54.1^\circ]$. Future experiments involving these two observables also have the potential to probe our model and to possibly further reduce the allowed parameter space. For instance, our 2\,$\sigma$ HPD results for $m_{\beta\beta}$, the effective mass parameter for the neutrinoless double beta decay, is predicted to be $m_{\beta\beta}\in[1.46, 2.24]$\,meV, well below the current experimental bound of $m_{\beta\beta}<61$\,meV provided by Ref.\ \cite{KamLAND-Zen:2016pfg}.

\section{Conclusions}
\label{sec:conclusions}
We present a minimal model of unification based on an $SU(5)$ gauge group augmented with a Peccei-Quinn symmetry that predicts the existence of ultralight axion dark matter within a narrow mass range of $m_a\in[0.1,\,4.7]$\,neV. This mass window is determined through an interplay between gauge coupling unification constraints, partial proton decay lifetime limits, and the need to reproduce the experimentally observed fermion mass spectrum. The model also predicts that neutrinos are purely of Majorana nature, possessing a normal mass hierarchy spectrum, where one of the neutrinos is a massless particle. We discuss the gauge boson mediated proton decay signatures of the model and specify expected partial lifetime ranges for two-body nucleon decays. Our analysis yields viable 2\,$\sigma$ ranges for the Dirac CP phase $\delta^\nu\in[-50.7^\circ, 55.6^\circ]$ and   for the neutrinoless double beta decay $m_{\beta\beta}\in[1.46, 2.24]$\,meV, respectively, through which the model may be tested in the neutrino experiments. 
Finally, we demonstrate that the entire parameter space of the model will be tested through a synergy between several low-energy experiments that look for proton decay (Hyper-Kamiokande) 
and axion dark matter (ABRACADABRA and DMRadio-GUT by measuring the axion-photon coupling, and CASPEr Electric by measuring the nucleon electric dipole moments).

\appendix
\section{Renormalization group running of the gauge couplings}\label{app:RGE}
The 2-loop renormalization group equations of the SM gauge couplings are given in Eq.~\ref{eq:2loop_gauge_running}. Here, we present the 1-loop and 2-loop gauge coefficients of  multiplets listed in Table~\ref{table:content}. The 1-loop gauge coefficients $\begin{pmatrix}b_1 & b_2 & b_3\end{pmatrix}$ are 
\begin{align}
    &b_i^{\phi_1^\textrm{Re}}=\begin{pmatrix}
        0 & \frac{1}{3} & 0
    \end{pmatrix},
    &&b_i^{\phi_1^\textrm{Im}}=\begin{pmatrix}
        0 & \frac{1}{3} & 0
    \end{pmatrix},
    &&b_i^{\phi_3^\textrm{Im}}=\begin{pmatrix}
        \frac{5}{12} & \frac{1}{4} & \frac{1}{6}
    \end{pmatrix},
    &&b_i^{\phi_{\overline{3}}^\textrm{Im}}=\begin{pmatrix}
        \frac{5}{12} & \frac{1}{4} & \frac{1}{6}
    \end{pmatrix},
    \nonumber\\ &
    b_i^{\phi_8^\textrm{Re}}=\begin{pmatrix}
        0 & 0 & \frac{1}{2}
    \end{pmatrix},
    &&b_i^{\phi_8^\textrm{Im}}=\begin{pmatrix}
        0 & 0 & \frac{1}{2}
    \end{pmatrix},
    &&b_i^{T_1}=\begin{pmatrix}
        \frac{1}{15} & 0 & \frac{1}{6}
    \end{pmatrix},
    &&b_i^{T_2}=\begin{pmatrix}
        \frac{1}{15} & 0 & \frac{1}{6}
    \end{pmatrix},
    \nonumber\\ &
    b_i^{\Phi_1}=\begin{pmatrix}
        \frac{9}{5} & \frac{5}{3} & 0
    \end{pmatrix},
    &&b_i^{\Phi_3}=\begin{pmatrix}
        \frac{4}{5} & 2 & \frac{1}{2}
    \end{pmatrix},
    &&b_i^{\Phi_6}=\begin{pmatrix}
        \frac{1}{15} & 1 & \frac{5}{3}
    \end{pmatrix},
    &&b_i^{\Phi_{10}}=\begin{pmatrix}
        2 & 0 & \frac{5}{2}
    \end{pmatrix},
    \nonumber\\ &
    b_i^{\Sigma_1}=\begin{pmatrix}
        \frac{6}{5} & \frac{4}{3} & 0
    \end{pmatrix},
    &&b_i^{\overline{\Sigma}_1}=\begin{pmatrix}
        \frac{6}{5} & \frac{4}{3} & 0
    \end{pmatrix},
    &&b_i^{\Sigma_3}=\begin{pmatrix}
        \frac{1}{15} & 1 & \frac{2}{3}
    \end{pmatrix},
    &&b_i^{\overline{\Sigma}_3}=\begin{pmatrix}
        \frac{1}{15} & 1 & \frac{2}{3}
    \end{pmatrix}, 
    \nonumber \\ &
    b_i^{\Sigma_6}=\begin{pmatrix}
        \frac{16}{15} & 0 & \frac{5}{3}
    \end{pmatrix},
    &&b_i^{\overline{\Sigma}_6}=\begin{pmatrix}
        \frac{16}{15} & 0 & \frac{5}{3}
    \end{pmatrix},
     &&b_i^{H_2}=\begin{pmatrix}
        \frac{1}{10} & \frac{1}{6} & 0
    \end{pmatrix},
\end{align}
whereas the 2-loop gauge coefficients read
\begin{alignat}{4}
    &b_{ij}^{\phi_1^\textrm{Re}}=\begin{pmatrix}
        0 & 0 & 0 \\
        0 & \frac{28}{3} & 0 \\
        0 & 0 & 0
    \end{pmatrix},
    &&b_{ij}^{\phi_1^\textrm{Im}}=\begin{pmatrix}
        0 & 0 & 0 \\
        0 & \frac{28}{3} & 0 \\
        0 & 0 & 0
    \end{pmatrix},
    &&b_{ij}^{\phi_3^\textrm{Im}}=\begin{pmatrix}
        \frac{25}{12} & \frac{15}{4} & \frac{20}{3} \\
        \frac{5}{4} & \frac{13}{4} & 4 \\
        \frac{5}{6} & \frac{3}{2} & \frac{11}{3}
    \end{pmatrix},
    &&b_{ij}^{\phi_{\overline{3}}^\textrm{Im}}=\begin{pmatrix}
        \frac{25}{12} & \frac{15}{4} & \frac{20}{3} \\
        \frac{5}{4} & \frac{13}{4} & 4 \\
        \frac{5}{6} & \frac{3}{2} & \frac{11}{3}
    \end{pmatrix},
    \nonumber\\
     &b_{ij}^{\phi_8^\textrm{Re}}=\begin{pmatrix}
        0 & 0 & 0 \\
        0 & 0 & 0 \\
        0 & 0 & 21
    \end{pmatrix},
     &&\;b_{ij}^{\phi_8^\textrm{Im}}=\begin{pmatrix}
        0 & 0 & 0 \\
        0 & 0 & 0 \\
        0 & 0 & 21
    \end{pmatrix},
    &&\;b_{ij}^{T_1}=\begin{pmatrix}
        \frac{4}{75} & 0 & \frac{16}{15} \\
        0 & 0 & 0 \\
        \frac{2}{15} & 0 & \frac{11}{3}
    \end{pmatrix},
    &&\;b_{ij}^{T_2}=\begin{pmatrix}
        \frac{4}{75} & 0 & \frac{16}{15} \\
        0 & 0 & 0 \\
        \frac{2}{15} & 0 & \frac{11}{3}
    \end{pmatrix},
    \nonumber\\
    &b_{ij}^{\Phi_1}=\begin{pmatrix}
        \frac{729}{25} & 81 & 0 \\
        27 & \frac{245}{3} & 0 \\
        0 & 0 & 0
    \end{pmatrix},
    &&\;b_{ij}^{\Phi_3}=\begin{pmatrix}
        \frac{64}{25} & \frac{96}{5} & \frac{64}{5} \\
        \frac{32}{5} & 56 & 32 \\
        \frac{8}{5} & 12 & 11
    \end{pmatrix},
    &&\;b_{ij}^{\Phi_6}=\begin{pmatrix}
        \frac{1}{75} & \frac{3}{5} & \frac{8}{3} \\
        \frac{1}{5} & 13 & 40 \\
        \frac{1}{3} & 15 & \frac{230}{3}
    \end{pmatrix},
    &&\;b_{ij}^{\Phi_{10}}=\begin{pmatrix}
        \frac{72}{5} & 0 & 144 \\
        0 & 0 & 0 \\
        18 & 0 & 195
    \end{pmatrix},
    \nonumber\\
    &b_{ij}^{\Sigma_{1}}=\begin{pmatrix}
        \frac{54}{25} & \frac{36}{5} & 0 \\
        \frac{12}{5} & \frac{64}{3} & 0 \\
        0 & 0 & 0
    \end{pmatrix},
    &&\;b_{ij}^{\overline{\Sigma}_{1}}=\begin{pmatrix}
        \frac{54}{25} & \frac{36}{5} & 0 \\
        \frac{12}{5} & \frac{64}{3} & 0 \\
        0 & 0 & 0
    \end{pmatrix},
    &&\;b_{ij}^{\Sigma_{3}}=\begin{pmatrix}
        \frac{1}{300} & \frac{3}{20} & \frac{4}{15} \\
        \frac{1}{20} & \frac{49}{4} & 4 \\
        \frac{1}{30} & \frac{3}{2} & \frac{38}{3}
    \end{pmatrix},
    &&\;b_{ij}^{\overline{\Sigma}_{3}}=\begin{pmatrix}
        \frac{1}{300} & \frac{3}{20} & \frac{4}{15} \\
        \frac{1}{20} & \frac{49}{4} & 4 \\
        \frac{1}{30} & \frac{3}{2} & \frac{38}{3}
    \end{pmatrix},
    \nonumber\\
    &b_{ij}^{\Sigma_{6}}=\begin{pmatrix}
        \frac{64}{75} & 0 & \frac{32}{3} \\
        0 & 0 & 0 \\
        \frac{4}{3} & 0 & \frac{125}{3}
    \end{pmatrix},
    &&\;b_{ij}^{\overline{\Sigma}_{6}}=\begin{pmatrix}
        \frac{64}{75} & 0 & \frac{32}{3} \\
        0 & 0 & 0 \\
        \frac{4}{3} & 0 & \frac{125}{3}
    \end{pmatrix},
    &&\;b_{ij}^{H_2}=\begin{pmatrix}
        \frac{9}{50} & \frac{9}{10} & 0 \\
        \frac{3}{10} & \frac{13}{6} & 0 \\
        0 & 0 & 0
    \end{pmatrix}.
\end{alignat}

\bibliographystyle{style}
\bibliography{reference}

\providecommand{\href}[2]{#2}\begingroup\raggedright\begin{thebibliography}{10}

\bibitem{Pati:1973rp}
J.~C. Pati and A.~Salam, ``{Is Baryon Number Conserved?},''
  \href{http://dx.doi.org/10.1103/PhysRevLett.31.661}{{\em Phys. Rev. Lett.}
  {\bfseries 31} (1973) 661--664}.

\bibitem{Pati:1974yy}
J.~C. Pati and A.~Salam, ``{Lepton Number as the Fourth Color},''
  \href{http://dx.doi.org/10.1103/PhysRevD.10.275}{{\em Phys. Rev. D}
  {\bfseries 10} (1974) 275--289}. [Erratum: Phys.Rev.D 11, 703--703 (1975)].

\bibitem{Georgi:1974sy}
H.~Georgi and S.~L. Glashow, ``{Unity of All Elementary Particle Forces},''
\href{http://dx.doi.org/10.1103/PhysRevLett.32.438}{{\em Phys. Rev. Lett.}
  {\bfseries 32} (1974) 438--441}.

\bibitem{Georgi:1974yf}
H.~Georgi, H.~R. Quinn, and S.~Weinberg, ``{Hierarchy of Interactions in
  Unified Gauge Theories},''
  \href{http://dx.doi.org/10.1103/PhysRevLett.33.451}{{\em Phys. Rev. Lett.}
  {\bfseries 33} (1974) 451--454}.

\bibitem{Georgi:1974my}
H.~Georgi, ``{The State of the Art---Gauge Theories},''
  \href{http://dx.doi.org/10.1063/1.2947450}{{\em AIP Conf. Proc.} {\bfseries
  23} (1975) 575--582}.

\bibitem{Fritzsch:1974nn}
H.~Fritzsch and P.~Minkowski, ``{Unified Interactions of Leptons and
  Hadrons},'' \href{http://dx.doi.org/10.1016/0003-4916(75)90211-0}{{\em Annals
  Phys.} {\bfseries 93} (1975) 193--266}.

\bibitem{Peccei:1977hh}
R.~D. Peccei and H.~R. Quinn, ``{CP Conservation in the Presence of
  Instantons},'' \href{http://dx.doi.org/10.1103/PhysRevLett.38.1440}{{\em
  Phys. Rev. Lett.} {\bfseries 38} (1977) 1440--1443}.

\bibitem{Peccei:1977ur}
R.~D. Peccei and H.~R. Quinn, ``{Constraints Imposed by CP Conservation in the
  Presence of Instantons},''
  \href{http://dx.doi.org/10.1103/PhysRevD.16.1791}{{\em Phys. Rev. D}
  {\bfseries 16} (1977) 1791--1797}.

\bibitem{Weinberg:1977ma}
S.~Weinberg, ``{A New Light Boson?},''
  \href{http://dx.doi.org/10.1103/PhysRevLett.40.223}{{\em Phys. Rev. Lett.}
  {\bfseries 40} (1978) 223--226}.

\bibitem{Wilczek:1977pj}
F.~Wilczek, ``{Problem of Strong $P$ and $T$ Invariance in the Presence of
  Instantons},'' \href{http://dx.doi.org/10.1103/PhysRevLett.40.279}{{\em Phys.
  Rev. Lett.} {\bfseries 40} (1978) 279--282}.

\bibitem{Kim:1979if}
J.~E. Kim, ``{Weak Interaction Singlet and Strong CP Invariance},''
  \href{http://dx.doi.org/10.1103/PhysRevLett.43.103}{{\em Phys. Rev. Lett.}
  {\bfseries 43} (1979) 103}.

\bibitem{Shifman:1979if}
M.~A. Shifman, A.~I. Vainshtein, and V.~I. Zakharov, ``{Can Confinement Ensure
  Natural CP Invariance of Strong Interactions?},''
  \href{http://dx.doi.org/10.1016/0550-3213(80)90209-6}{{\em Nucl. Phys. B}
  {\bfseries 166} (1980) 493--506}.

\bibitem{Zhitnitsky:1980tq}
A.~R. Zhitnitsky, ``{On Possible Suppression of the Axion Hadron Interactions.
  (In Russian)},'' {\em Sov. J. Nucl. Phys.} {\bfseries 31} (1980) 260.

\bibitem{Dine:1981rt}
M.~Dine, W.~Fischler, and M.~Srednicki, ``{A Simple Solution to the Strong CP
  Problem with a Harmless Axion},''
  \href{http://dx.doi.org/10.1016/0370-2693(81)90590-6}{{\em Phys. Lett. B}
  {\bfseries 104} (1981) 199--202}.

\bibitem{Preskill:1982cy}
J.~Preskill, M.~B. Wise, and F.~Wilczek, ``{Cosmology of the Invisible
  Axion},'' \href{http://dx.doi.org/10.1016/0370-2693(83)90637-8}{{\em Phys.
  Lett. B} {\bfseries 120} (1983) 127--132}.

\bibitem{Abbott:1982af}
L.~F. Abbott and P.~Sikivie, ``{A Cosmological Bound on the Invisible Axion},''
  \href{http://dx.doi.org/10.1016/0370-2693(83)90638-X}{{\em Phys. Lett. B}
  {\bfseries 120} (1983) 133--136}.

\bibitem{Dine:1982ah}
M.~Dine and W.~Fischler, ``{The Not So Harmless Axion},''
  \href{http://dx.doi.org/10.1016/0370-2693(83)90639-1}{{\em Phys. Lett. B}
  {\bfseries 120} (1983) 137--141}.

\bibitem{Wise:1981ry}
M.~B. Wise, H.~Georgi, and S.~L. Glashow, ``{SU(5) and the Invisible Axion},''
  \href{http://dx.doi.org/10.1103/PhysRevLett.47.402}{{\em Phys. Rev. Lett.}
  {\bfseries 47} (1981) 402}.

\bibitem{Ernst:2018bib}
A.~Ernst, A.~Ringwald, and C.~Tamarit, ``{Axion Predictions in $SO(10)\times
  U(1)_{\rm PQ}$ Models},''
  \href{http://dx.doi.org/10.1007/JHEP02(2018)103}{{\em JHEP} {\bfseries 02}
  (2018) 103}, \href{http://arxiv.org/abs/1801.04906}{{\ttfamily
  arXiv:1801.04906 [hep-ph]}}.

\bibitem{DiLuzio:2018gqe}
L.~Di~Luzio, A.~Ringwald, and C.~Tamarit, ``{Axion mass prediction from minimal
  grand unification},''
  \href{http://dx.doi.org/10.1103/PhysRevD.98.095011}{{\em Phys. Rev. D}
  {\bfseries 98} no.~9, (2018) 095011},
  \href{http://arxiv.org/abs/1807.09769}{{\ttfamily arXiv:1807.09769
  [hep-ph]}}.

\bibitem{FileviezPerez:2019fku}
P.~Fileviez~P\'erez, C.~Murgui, and A.~D. Plascencia, ``{The QCD Axion and
  Unification},'' \href{http://dx.doi.org/10.1007/JHEP11(2019)093}{{\em JHEP}
  {\bfseries 11} (2019) 093}, \href{http://arxiv.org/abs/1908.01772}{{\ttfamily
  arXiv:1908.01772 [hep-ph]}}.

\bibitem{FileviezPerez:2019ssf}
P.~Fileviez~P\'erez, C.~Murgui, and A.~D. Plascencia, ``{Axion Dark Matter,
  Proton Decay and Unification},''
  \href{http://dx.doi.org/10.1007/JHEP01(2020)091}{{\em JHEP} {\bfseries 01}
  (2020) 091}, \href{http://arxiv.org/abs/1911.05738}{{\ttfamily
  arXiv:1911.05738 [hep-ph]}}.

\bibitem{Agrawal:2022lsp}
P.~Agrawal, M.~Nee, and M.~Reig, ``{Axion couplings in grand unified
  theories},'' \href{http://dx.doi.org/10.1007/JHEP10(2022)141}{{\em JHEP}
  {\bfseries 10} (2022) 141}, \href{http://arxiv.org/abs/2206.07053}{{\ttfamily
  arXiv:2206.07053 [hep-ph]}}.

\bibitem{Dorsner:2019vgf}
I.~Dor\v{s}ner and S.~Saad, ``{Towards Minimal $SU(5)$},''
  \href{http://dx.doi.org/10.1103/PhysRevD.101.015009}{{\em Phys. Rev.}
  {\bfseries D101} no.~1, (2020) 015009},
\href{http://arxiv.org/abs/1910.09008}{{\ttfamily arXiv:1910.09008 [hep-ph]}}.

\bibitem{Dorsner:2021qwg}
I.~Dor\v{s}ner, E.~D\v{z}aferovi\'c-Ma\v{s}i\'c, and S.~Saad, ``{Parameter
  space exploration of the minimal SU(5) unification},''
  \href{http://dx.doi.org/10.1103/PhysRevD.104.015023}{{\em Phys. Rev. D}
  {\bfseries 104} no.~1, (2021) 015023},
  \href{http://arxiv.org/abs/2105.01678}{{\ttfamily arXiv:2105.01678
  [hep-ph]}}.

\bibitem{Oshimo:2009ia}
N.~Oshimo, ``{Realistic model for SU(5) grand unification},''
  \href{http://dx.doi.org/10.1103/PhysRevD.80.075011}{{\em Phys. Rev.}
  {\bfseries D80} (2009) 075011},
\href{http://arxiv.org/abs/0907.3400}{{\ttfamily arXiv:0907.3400 [hep-ph]}}.

\bibitem{Babu:2009aq}
K.~S. Babu, S.~Nandi, and Z.~Tavartkiladze, ``{New Mechanism for Neutrino Mass
  Generation and Triply Charged Higgs Bosons at the LHC},''
  \href{http://dx.doi.org/10.1103/PhysRevD.80.071702}{{\em Phys. Rev.}
  {\bfseries D80} (2009) 071702},
\href{http://arxiv.org/abs/0905.2710}{{\ttfamily arXiv:0905.2710 [hep-ph]}}.

\bibitem{Bambhaniya:2013yca}
G.~Bambhaniya, J.~Chakrabortty, S.~Goswami, and P.~Konar, ``{Generation of
  neutrino mass from new physics at TeV scale and multilepton signatures at the
  LHC},'' \href{http://dx.doi.org/10.1103/PhysRevD.88.075006}{{\em Phys. Rev.
  D} {\bfseries 88} no.~7, (2013) 075006},
  \href{http://arxiv.org/abs/1305.2795}{{\ttfamily arXiv:1305.2795 [hep-ph]}}.

\bibitem{ParticleDataGroup:2022pth}
{\bfseries Particle Data Group} Collaboration, R.~L. Workman {\em et~al.},
  ``{Review of Particle Physics},''
  \href{http://dx.doi.org/10.1093/ptep/ptac097}{{\em PTEP} {\bfseries 2022}
  (2022) 083C01}.

\bibitem{Cordero-Carrion:2018xre}
I.~Cordero-Carri\'on, M.~Hirsch, and A.~Vicente, ``{Master Majorana neutrino
  mass parametrization},''
  \href{http://dx.doi.org/10.1103/PhysRevD.99.075019}{{\em Phys. Rev. D}
  {\bfseries 99} no.~7, (2019) 075019},
  \href{http://arxiv.org/abs/1812.03896}{{\ttfamily arXiv:1812.03896
  [hep-ph]}}.

\bibitem{Cordero-Carrion:2019qtu}
I.~Cordero-Carri\'on, M.~Hirsch, and A.~Vicente, ``{General parametrization of
  Majorana neutrino mass models},''
  \href{http://dx.doi.org/10.1103/PhysRevD.101.075032}{{\em Phys. Rev. D}
  {\bfseries 101} no.~7, (2020) 075032},
  \href{http://arxiv.org/abs/1912.08858}{{\ttfamily arXiv:1912.08858
  [hep-ph]}}.

\bibitem{Dorsner:2005fq}
I.~Dorsner and P.~Fileviez~Perez, ``{Unification without supersymmetry:
  Neutrino mass, proton decay and light leptoquarks},''
  \href{http://dx.doi.org/10.1016/j.nuclphysb.2005.06.016}{{\em Nucl. Phys.}
  {\bfseries B723} (2005) 53--76},
\href{http://arxiv.org/abs/hep-ph/0504276}{{\ttfamily arXiv:hep-ph/0504276
  [hep-ph]}}.

\bibitem{Bajc:2006ia}
B.~Bajc and G.~Senjanovic, ``{Seesaw at LHC},''
  \href{http://dx.doi.org/10.1088/1126-6708/2007/08/014}{{\em JHEP} {\bfseries
  08} (2007) 014},
\href{http://arxiv.org/abs/hep-ph/0612029}{{\ttfamily arXiv:hep-ph/0612029
  [hep-ph]}}.

\bibitem{Perez:2007rm}
P.~Fileviez~Perez, ``{Renormalizable adjoint SU(5)},''
  \href{http://dx.doi.org/10.1016/j.physletb.2007.07.075}{{\em Phys. Lett.}
  {\bfseries B654} (2007) 189--193},
\href{http://arxiv.org/abs/hep-ph/0702287}{{\ttfamily arXiv:hep-ph/0702287
  [hep-ph]}}.

\bibitem{Perez:2016qbo}
P.~Fileviez~Perez and C.~Murgui, ``{Renormalizable SU(5) Unification},''
  \href{http://dx.doi.org/10.1103/PhysRevD.94.075014}{{\em Phys. Rev.}
  {\bfseries D94} no.~7, (2016) 075014},
\href{http://arxiv.org/abs/1604.03377}{{\ttfamily arXiv:1604.03377 [hep-ph]}}.

\bibitem{Dorsner:2017wwn}
I.~Doršner, S.~Fajfer, and N.~Košnik, ``{Leptoquark mechanism of neutrino
  masses within the grand unification framework},''
  \href{http://dx.doi.org/10.1140/epjc/s10052-017-4987-2}{{\em Eur. Phys. J.}
  {\bfseries C77} no.~6, (2017) 417},
\href{http://arxiv.org/abs/1701.08322}{{\ttfamily arXiv:1701.08322 [hep-ph]}}.

\bibitem{Kumericki:2017sfc}
K.~Kumericki, T.~Mede, and I.~Picek, ``{Renormalizable SU(5) Completions of a
  Zee-type Neutrino Mass Model},''
  \href{http://dx.doi.org/10.1103/PhysRevD.97.055012}{{\em Phys. Rev.}
  {\bfseries D97} no.~5, (2018) 055012},
\href{http://arxiv.org/abs/1712.05246}{{\ttfamily arXiv:1712.05246 [hep-ph]}}.

\bibitem{Saad:2019vjo}
S.~Saad, ``{Origin of a two-loop neutrino mass from SU(5) grand unification},''
  \href{http://dx.doi.org/10.1103/PhysRevD.99.115016}{{\em Phys. Rev.}
  {\bfseries D99} no.~11, (2019) 115016},
\href{http://arxiv.org/abs/1902.11254}{{\ttfamily arXiv:1902.11254 [hep-ph]}}.

\bibitem{Klein:2019jgb}
C.~Klein, M.~Lindner, and S.~Vogl, ``{Radiative neutrino masses and successful
  $SU(5)$ unification},''
  \href{http://dx.doi.org/10.1103/PhysRevD.100.075024}{{\em Phys. Rev.}
  {\bfseries D100} no.~7, (2019) 075024},
\href{http://arxiv.org/abs/1907.05328}{{\ttfamily arXiv:1907.05328 [hep-ph]}}.

\bibitem{Antusch:2021yqe}
S.~Antusch and K.~Hinze, ``{Nucleon decay in a minimal non-SUSY GUT with
  predicted quark-lepton Yukawa ratios},''
  \href{http://dx.doi.org/10.1016/j.nuclphysb.2022.115719}{{\em Nucl. Phys. B}
  {\bfseries 976} (2022) 115719},
  \href{http://arxiv.org/abs/2108.08080}{{\ttfamily arXiv:2108.08080
  [hep-ph]}}.

\bibitem{Antusch:2022afk}
S.~Antusch, K.~Hinze, and S.~Saad, ``{Viable quark-lepton Yukawa ratios and
  nucleon decay predictions in $SU(5)$ GUTs with type-II seesaw},''
  \href{http://arxiv.org/abs/2205.01120}{{\ttfamily arXiv:2205.01120
  [hep-ph]}}.

\bibitem{Srednicki:1985xd}
M.~Srednicki, ``{Axion Couplings to Matter. 1. CP Conserving Parts},''
  \href{http://dx.doi.org/10.1016/0550-3213(85)90054-9}{{\em Nucl. Phys. B}
  {\bfseries 260} (1985) 689--700}.

\bibitem{DiLuzio:2020wdo}
L.~Di~Luzio, M.~Giannotti, E.~Nardi, and L.~Visinelli, ``{The landscape of QCD
  axion models},'' \href{http://dx.doi.org/10.1016/j.physrep.2020.06.002}{{\em
  Phys. Rept.} {\bfseries 870} (2020) 1--117},
  \href{http://arxiv.org/abs/2003.01100}{{\ttfamily arXiv:2003.01100
  [hep-ph]}}.

\bibitem{Bardeen:1978nq}
W.~A. Bardeen, S.~H.~H. Tye, and J.~A.~M. Vermaseren, ``{Phenomenology of the
  New Light Higgs Boson Search},''
  \href{http://dx.doi.org/10.1016/0370-2693(78)90859-6}{{\em Phys. Lett. B}
  {\bfseries 76} (1978) 580--584}.

\bibitem{GrillidiCortona:2015jxo}
G.~Grilli~di Cortona, E.~Hardy, J.~Pardo~Vega, and G.~Villadoro, ``{The QCD
  axion, precisely},'' \href{http://dx.doi.org/10.1007/JHEP01(2016)034}{{\em
  JHEP} {\bfseries 01} (2016) 034},
  \href{http://arxiv.org/abs/1511.02867}{{\ttfamily arXiv:1511.02867
  [hep-ph]}}.

\bibitem{Kahn:2016aff}
Y.~Kahn, B.~R. Safdi, and J.~Thaler, ``{Broadband and Resonant Approaches to
  Axion Dark Matter Detection},''
  \href{http://dx.doi.org/10.1103/PhysRevLett.117.141801}{{\em Phys. Rev.
  Lett.} {\bfseries 117} no.~14, (2016) 141801},
  \href{http://arxiv.org/abs/1602.01086}{{\ttfamily arXiv:1602.01086
  [hep-ph]}}.

\bibitem{Domcke:2022rgu}
V.~Domcke, C.~Garcia-Cely, and N.~L. Rodd, ``{Novel Search for High-Frequency
  Gravitational Waves with Low-Mass Axion Haloscopes},''
  \href{http://dx.doi.org/10.1103/PhysRevLett.129.041101}{{\em Phys. Rev.
  Lett.} {\bfseries 129} no.~4, (2022) 041101},
  \href{http://arxiv.org/abs/2202.00695}{{\ttfamily arXiv:2202.00695
  [hep-ph]}}.

\bibitem{DMRadio}
``Dmradio-gut: Probing gut-scale qcd axion dark matter,
  \url{https://www.snowmass21.org/docs/files/summaries/CF/SNOWMASS21-CF2_CF0-IF1_IF0_Saptarshi_Chaudhuri-219.pdf},''.

\bibitem{Murgui:2022zvy}
C.~Murgui, Y.~Wang, and K.~M. Zurek, ``{Axion Detection with Optomechanical
  Cavities},'' \href{http://arxiv.org/abs/2211.08432}{{\ttfamily
  arXiv:2211.08432 [hep-ph]}}.

\bibitem{Adams:2022pbo}
C.~B. Adams {\em et~al.}, ``{Axion Dark Matter},'' in {\em {2022 Snowmass
  Summer Study}}.
\newblock 3, 2022.
\newblock \href{http://arxiv.org/abs/2203.14923}{{\ttfamily arXiv:2203.14923
  [hep-ex]}}.

\bibitem{Budker:2013hfa}
D.~Budker, P.~W. Graham, M.~Ledbetter, S.~Rajendran, and A.~Sushkov,
  ``{Proposal for a Cosmic Axion Spin Precession Experiment (CASPEr)},''
  \href{http://dx.doi.org/10.1103/PhysRevX.4.021030}{{\em Phys. Rev. X}
  {\bfseries 4} no.~2, (2014) 021030},
  \href{http://arxiv.org/abs/1306.6089}{{\ttfamily arXiv:1306.6089 [hep-ph]}}.

\bibitem{JacksonKimball:2017elr}
D.~F. Jackson~Kimball {\em et~al.}, ``{Overview of the Cosmic Axion Spin
  Precession Experiment (CASPEr)},''
  \href{http://dx.doi.org/10.1007/978-3-030-43761-9_13}{{\em Springer Proc.
  Phys.} {\bfseries 245} (2020) 105--121},
  \href{http://arxiv.org/abs/1711.08999}{{\ttfamily arXiv:1711.08999
  [physics.ins-det]}}.

\bibitem{Graham:2013gfa}
P.~W. Graham and S.~Rajendran, ``{New Observables for Direct Detection of Axion
  Dark Matter},'' \href{http://dx.doi.org/10.1103/PhysRevD.88.035023}{{\em
  Phys. Rev. D} {\bfseries 88} (2013) 035023},
  \href{http://arxiv.org/abs/1306.6088}{{\ttfamily arXiv:1306.6088 [hep-ph]}}.

\bibitem{Pospelov:1999ha}
M.~Pospelov and A.~Ritz, ``{Theta induced electric dipole moment of the neutron
  via QCD sum rules},''
  \href{http://dx.doi.org/10.1103/PhysRevLett.83.2526}{{\em Phys. Rev. Lett.}
  {\bfseries 83} (1999) 2526--2529},
  \href{http://arxiv.org/abs/hep-ph/9904483}{{\ttfamily arXiv:hep-ph/9904483}}.

\bibitem{Crewther:1979pi}
R.~J. Crewther, P.~Di~Vecchia, G.~Veneziano, and E.~Witten, ``{Chiral Estimate
  of the Electric Dipole Moment of the Neutron in Quantum Chromodynamics},''
  \href{http://dx.doi.org/10.1016/0370-2693(79)90128-X}{{\em Phys. Lett. B}
  {\bfseries 88} (1979) 123}. [Erratum: Phys.Lett.B 91, 487 (1980)].

\bibitem{Hisano:2012sc}
J.~Hisano, J.~Y. Lee, N.~Nagata, and Y.~Shimizu, ``{Reevaluation of Neutron
  Electric Dipole Moment with QCD Sum Rules},''
  \href{http://dx.doi.org/10.1103/PhysRevD.85.114044}{{\em Phys. Rev. D}
  {\bfseries 85} (2012) 114044},
  \href{http://arxiv.org/abs/1204.2653}{{\ttfamily arXiv:1204.2653 [hep-ph]}}.

\bibitem{Yoon:2017tag}
B.~Yoon, T.~Bhattacharya, and R.~Gupta, ``{Neutron Electric Dipole Moment on
  the Lattice},'' \href{http://dx.doi.org/10.1051/epjconf/201817501014}{{\em
  EPJ Web Conf.} {\bfseries 175} (2018) 01014},
  \href{http://arxiv.org/abs/1712.08557}{{\ttfamily arXiv:1712.08557
  [hep-lat]}}.

\bibitem{Ballesteros:2016xej}
G.~Ballesteros, J.~Redondo, A.~Ringwald, and C.~Tamarit, ``{Standard
  Model\textemdash{}axion\textemdash{}seesaw\textemdash{}Higgs portal
  inflation. Five problems of particle physics and cosmology solved in one
  stroke},'' \href{http://dx.doi.org/10.1088/1475-7516/2017/08/001}{{\em JCAP}
  {\bfseries 08} (2017) 001}, \href{http://arxiv.org/abs/1610.01639}{{\ttfamily
  arXiv:1610.01639 [hep-ph]}}.

\bibitem{Planck:2018vyg}
{\bfseries Planck} Collaboration, N.~Aghanim {\em et~al.}, ``{Planck 2018
  results. VI. Cosmological parameters},''
  \href{http://dx.doi.org/10.1051/0004-6361/201833910}{{\em Astron. Astrophys.}
  {\bfseries 641} (2020) A6}, \href{http://arxiv.org/abs/1807.06209}{{\ttfamily
  arXiv:1807.06209 [astro-ph.CO]}}. [Erratum: Astron.Astrophys. 652, C4
  (2021)].

\bibitem{Machacek:1983tz}
M.~E. Machacek and M.~T. Vaughn, ``{Two Loop Renormalization Group Equations in
  a General Quantum Field Theory. 1. Wave Function Renormalization},''
  \href{http://dx.doi.org/10.1016/0550-3213(83)90610-7}{{\em Nucl. Phys. B}
  {\bfseries 222} (1983) 83--103}.

\bibitem{Antusch:2013jca}
S.~Antusch and V.~Maurer, ``{Running quark and lepton parameters at various
  scales},'' \href{http://dx.doi.org/10.1007/JHEP11(2013)115}{{\em JHEP}
  {\bfseries 11} (2013) 115},
\href{http://arxiv.org/abs/1306.6879}{{\ttfamily arXiv:1306.6879 [hep-ph]}}.

\bibitem{Claudson:1981gh}
M.~Claudson, M.~B. Wise, and L.~J. Hall, ``{Chiral Lagrangian for Deep Mine
  Physics},'' \href{http://dx.doi.org/10.1016/0550-3213(82)90401-1}{{\em Nucl.
  Phys. B} {\bfseries 195} (1982) 297--307}.

\bibitem{JLQCD:1999dld}
{\bfseries JLQCD} Collaboration, S.~Aoki {\em et~al.}, ``{Nucleon decay matrix
  elements from lattice QCD},''
  \href{http://dx.doi.org/10.1103/PhysRevD.62.014506}{{\em Phys. Rev. D}
  {\bfseries 62} (2000) 014506},
  \href{http://arxiv.org/abs/hep-lat/9911026}{{\ttfamily
  arXiv:hep-lat/9911026}}.

\bibitem{Antusch:2020ztu}
S.~Antusch, C.~Hohl, and V.~Susi\v{c}, ``{Employing nucleon decay as a
  fingerprint of SUSY GUT models using SusyTCProton},''
  \href{http://dx.doi.org/10.1007/JHEP06(2021)022}{{\em JHEP} {\bfseries 06}
  (2021) 022}, \href{http://arxiv.org/abs/2011.15026}{{\ttfamily
  arXiv:2011.15026 [hep-ph]}}.

\bibitem{Nihei:1994tx}
T.~Nihei and J.~Arafune, ``{The Two loop long range effect on the proton decay
  effective Lagrangian},'' \href{http://dx.doi.org/10.1143/ptp/93.3.665,
  10.1143/PTP.93.665}{{\em Prog. Theor. Phys.} {\bfseries 93} (1995) 665--669},
\href{http://arxiv.org/abs/hep-ph/9412325}{{\ttfamily arXiv:hep-ph/9412325
  [hep-ph]}}.

\bibitem{Wilczek:1979hc}
F.~Wilczek and A.~Zee, ``{Operator Analysis of Nucleon Decay},''
\href{http://dx.doi.org/10.1103/PhysRevLett.43.1571}{{\em Phys. Rev. Lett.}
  {\bfseries 43} (1979) 1571--1573}.

\bibitem{Buras:1977yy}
A.~J. Buras, J.~R. Ellis, M.~K. Gaillard, and D.~V. Nanopoulos, ``{Aspects of
  the Grand Unification of Strong, Weak and Electromagnetic Interactions},''
\href{http://dx.doi.org/10.1016/0550-3213(78)90214-6}{{\em Nucl. Phys.}
  {\bfseries B135} (1978) 66--92}.

\bibitem{Ellis:1979hy}
J.~R. Ellis, M.~K. Gaillard, and D.~V. Nanopoulos, ``{On the Effective
  Lagrangian for Baryon Decay},''
\href{http://dx.doi.org/10.1016/0370-2693(79)90477-5}{{\em Phys. Lett.}
  {\bfseries 88B} (1979) 320--324}.

\bibitem{Aoki:2017puj}
Y.~Aoki, T.~Izubuchi, E.~Shintani, and A.~Soni, ``{Improved lattice computation
  of proton decay matrix elements},''
  \href{http://dx.doi.org/10.1103/PhysRevD.96.014506}{{\em Phys. Rev.}
  {\bfseries D96} no.~1, (2017) 014506},
\href{http://arxiv.org/abs/1705.01338}{{\ttfamily arXiv:1705.01338 [hep-lat]}}.

\bibitem{Yoo:2021gql}
J.-S. Yoo, Y.~Aoki, P.~Boyle, T.~Izubuchi, A.~Soni, and S.~Syritsyn, ``{Proton
  decay matrix elements on the lattice at physical pion mass},''
  \href{http://dx.doi.org/10.1103/PhysRevD.105.074501}{{\em Phys. Rev. D}
  {\bfseries 105} no.~7, (2022) 074501},
  \href{http://arxiv.org/abs/2111.01608}{{\ttfamily arXiv:2111.01608
  [hep-lat]}}.

\bibitem{DeRujula:1980qc}
A.~De~Rujula, H.~Georgi, and S.~L. Glashow, ``{FLAVOR GONIOMETRY BY PROTON
  DECAY},'' \href{http://dx.doi.org/10.1103/PhysRevLett.45.413}{{\em Phys. Rev.
  Lett.} {\bfseries 45} (1980) 413}.

\bibitem{FileviezPerez:2004hn}
P.~Fileviez~Perez, ``{Fermion mixings versus d = 6 proton decay},''
  \href{http://dx.doi.org/10.1016/j.physletb.2004.06.061}{{\em Phys. Lett.}
  {\bfseries B595} (2004) 476--483},
\href{http://arxiv.org/abs/hep-ph/0403286}{{\ttfamily arXiv:hep-ph/0403286
  [hep-ph]}}.

\bibitem{Nath:2006ut}
P.~Nath and P.~Fileviez~Perez, ``{Proton stability in grand unified theories,
  in strings and in branes},''
  \href{http://dx.doi.org/10.1016/j.physrep.2007.02.010}{{\em Phys. Rept.}
  {\bfseries 441} (2007) 191--317},
  \href{http://arxiv.org/abs/hep-ph/0601023}{{\ttfamily arXiv:hep-ph/0601023}}.

\bibitem{Dev:2022jbf}
P.~S.~B. Dev {\em et~al.}, ``{Searches for Baryon Number Violation in Neutrino
  Experiments: A White Paper},''
  \href{http://arxiv.org/abs/2203.08771}{{\ttfamily arXiv:2203.08771
  [hep-ex]}}.

\bibitem{Super-Kamiokande:2020wjk}
{\bfseries Super-Kamiokande} Collaboration, A.~Takenaka {\em et~al.}, ``{Search
  for proton decay via $p\to e^+\pi^0$ and $p\to \mu^+\pi^0$ with an enlarged
  fiducial volume in Super-Kamiokande I-IV},''
  \href{http://dx.doi.org/10.1103/PhysRevD.102.112011}{{\em Phys. Rev. D}
  {\bfseries 102} no.~11, (2020) 112011},
  \href{http://arxiv.org/abs/2010.16098}{{\ttfamily arXiv:2010.16098
  [hep-ex]}}.

\bibitem{Hyper-Kamiokande:2018ofw}
{\bfseries Hyper-Kamiokande} Collaboration, K.~Abe {\em et~al.},
  ``{Hyper-Kamiokande Design Report},''
  \href{http://arxiv.org/abs/1805.04163}{{\ttfamily arXiv:1805.04163
  [physics.ins-det]}}.

\bibitem{Super-Kamiokande:2017gev}
{\bfseries Super-Kamiokande} Collaboration, K.~Abe {\em et~al.}, ``{Search for
  nucleon decay into charged antilepton plus meson in 0.316 megaton$\cdot$years
  exposure of the Super-Kamiokande water Cherenkov detector},''
  \href{http://dx.doi.org/10.1103/PhysRevD.96.012003}{{\em Phys. Rev. D}
  {\bfseries 96} no.~1, (2017) 012003},
  \href{http://arxiv.org/abs/1705.07221}{{\ttfamily arXiv:1705.07221
  [hep-ex]}}.

\bibitem{Brock:2012ogj}
R.~Brock {\em et~al.}, ``{Proton Decay},'' in {\em {Workshop on Fundamental
  Physics at the Intensity Frontier}}, pp.~111--130.
\newblock 5, 2012.

\bibitem{Super-Kamiokande:2022egr}
{\bfseries Super-Kamiokande} Collaboration, R.~Matsumoto {\em et~al.},
  ``{Search for proton decay via $p\rightarrow \mu^+K^0$ in 0.37 megaton-years
  exposure of Super-Kamiokande},''
  \href{http://dx.doi.org/10.1103/PhysRevD.106.072003}{{\em Phys. Rev. D}
  {\bfseries 106} no.~7, (2022) 072003},
  \href{http://arxiv.org/abs/2208.13188}{{\ttfamily arXiv:2208.13188
  [hep-ex]}}.

\bibitem{Super-Kamiokande:2013rwg}
{\bfseries Super-Kamiokande} Collaboration, K.~Abe {\em et~al.}, ``{Search for
  Nucleon Decay via $n \to \bar{\nu} \pi^{0}$ and $p \to \bar{\nu} \pi^{+}$ in
  Super-Kamiokande},''
  \href{http://dx.doi.org/10.1103/PhysRevLett.113.121802}{{\em Phys. Rev.
  Lett.} {\bfseries 113} no.~12, (2014) 121802},
  \href{http://arxiv.org/abs/1305.4391}{{\ttfamily arXiv:1305.4391 [hep-ex]}}.

\bibitem{Takhistov:2016eqm}
{\bfseries Super-Kamiokande} Collaboration, V.~Takhistov, ``{Review of Nucleon
  Decay Searches at Super-Kamiokande},'' in {\em {51st Rencontres de Moriond on
  EW Interactions and Unified Theories}}, pp.~437--444.
\newblock 2016.
\newblock \href{http://arxiv.org/abs/1605.03235}{{\ttfamily arXiv:1605.03235
  [hep-ex]}}.

\bibitem{Babu:2016bmy}
K.~S. Babu, B.~Bajc, and S.~Saad, ``{Yukawa Sector of Minimal SO(10)
  Unification},'' \href{http://dx.doi.org/10.1007/JHEP02(2017)136}{{\em JHEP}
  {\bfseries 02} (2017) 136},
\href{http://arxiv.org/abs/1612.04329}{{\ttfamily arXiv:1612.04329 [hep-ph]}}.

\bibitem{Branco:2011iw}
G.~C. Branco, P.~M. Ferreira, L.~Lavoura, M.~N. Rebelo, M.~Sher, and J.~P.
  Silva, ``{Theory and phenomenology of two-Higgs-doublet models},''
  \href{http://dx.doi.org/10.1016/j.physrep.2012.02.002}{{\em Phys. Rept.}
  {\bfseries 516} (2012) 1--102},
  \href{http://arxiv.org/abs/1106.0034}{{\ttfamily arXiv:1106.0034 [hep-ph]}}.

\bibitem{Hoang:2020iah}
A.~H. Hoang, ``{What is the Top Quark Mass?},''
  \href{http://dx.doi.org/10.1146/annurev-nucl-101918-023530}{{\em Ann. Rev.
  Nucl. Part. Sci.} {\bfseries 70} (2020) 225--255},
  \href{http://arxiv.org/abs/2004.12915}{{\ttfamily arXiv:2004.12915
  [hep-ph]}}.

\bibitem{KamLAND-Zen:2016pfg}
{\bfseries KamLAND-Zen} Collaboration, A.~Gando {\em et~al.}, ``{Search for
  Majorana Neutrinos near the Inverted Mass Hierarchy Region with
  KamLAND-Zen},'' \href{http://dx.doi.org/10.1103/PhysRevLett.117.082503}{{\em
  Phys. Rev. Lett.} {\bfseries 117} no.~8, (2016) 082503},
  \href{http://arxiv.org/abs/1605.02889}{{\ttfamily arXiv:1605.02889
  [hep-ex]}}. [Addendum: Phys.Rev.Lett. 117, 109903 (2016)].

\end{thebibliography}\endgroup
\end{document}